\documentclass[aps,prd,eqsecnum,nofootinbib,superscriptaddress,11pt]{revtex4}
\usepackage{amssymb,amsmath,psfrag,epsfig}

\usepackage[bookmarks=true,hyperfigures=true]{hyperref}
\usepackage{graphicx}
\usepackage{multirow}

\usepackage{cancel}
\newcommand{\beq}{\begin{equation}}
\newcommand{\eeq}{\end{equation}}
\newcommand{\beqa}{\begin{eqnarray}}
\newcommand{\eeqa}{\end{eqnarray}}
\newcommand{\bea}{\begin{align}}
\newcommand{\eal}{\end{align}}
\newcommand{\nn}{\nonumber  \\}

\makeatletter
\newlength{\apb@width}
\newcommand{\autoparbox}[2][c]{\settowidth{\apb@width}{#2}\parbox[#1]{\apb@width}{#2}}
\newcommand{\includegraphicsbox}[2][]{\autoparbox{\includegraphics[#1]{#2}}}
\makeatother

\ifx\href\asklfhas\newcommand{\href}[2]{#2}\fi
\newcommand{\arxivlink}[1]{\href{http://arxiv.org/abs/#1}{arxiv:#1}}

\usepackage[font=small,labelfont=bf,width=0.85\textwidth]{caption}

\ifx\hypersetup\sadfkjashdfkxja\else \hypersetup{pdftitle={On
All-loop Integrands of Scattering Amplitudes of Planar N=4 SYM}}
\hypersetup{pdfauthor={Song He and Tristan McLoughlin}}
\hypersetup{pdfsubject={}} \hypersetup{pdfkeywords={Scattering
Amplitude, N=4 Supersymmetric Gauge Theory, All-loop}}
\hypersetup{plainpages=false} \hypersetup{pdfpagemode=UseNone}
\hypersetup{bookmarksnumbered=true} \hypersetup{pdfstartview=FitH}
\hypersetup{colorlinks=false} \hypersetup{citebordercolor={.5 1 .5}}
\hypersetup{urlbordercolor={.5 1 1}} \hypersetup{linkbordercolor={1
.7 .7}}
\fi

\begin{document}

\thispagestyle{empty}
\begin{flushright}\footnotesize
\texttt{\arxivlink{1010.6256}}\\
\texttt{AEI-2010-158}%
\end{flushright}
\vspace{1cm}

\begin{center}%
{\Large\textbf{\mathversion{bold}%
On All-loop Integrands of Scattering Amplitudes in Planar\\
 $\mathcal{N}=4$ SYM}\par}
\vspace{1cm}%

\textsc{Song He \& Tristan McLoughlin}\vspace{5mm}%

\textit{Max-Planck-Institut f\"ur Gravitationsphysik\\%
Albert-Einstein-Institut\\%
Am M\"uhlenberg 1, 14476 Potsdam,
\\
 Germany}\vspace{3mm}%

\{\texttt{song.he, tristan.mcloughlin}\}\texttt{@aei.mpg.de}
\par\vspace{1cm}

\vspace{1cm}

\textbf{Abstract}\vspace{7mm}

\begin{minipage}{12.7cm}
We study the relationship between the momentum twistor MHV vertex
expansion of planar amplitudes in ${\cal N}=4$ super-Yang--Mills and
the all-loop generalization of the BCFW recursion relations. We
demonstrate explicitly in several examples that the MHV vertex expressions
for tree-level
amplitudes and loop integrands  satisfy
the recursion relations. Furthermore, we introduce a rewriting of
the MHV expansion in terms of sums over non-crossing partitions and
show that this cyclically invariant formula satisfies the recursion
relations for all numbers of legs and all loop orders.
\end{minipage}

\end{center}

\newpage

\section{Introduction}

The MHV vertex expansion, due to Cachazo et al (CSW)
\cite{Cachazo:2004kj}, is a diagrammatic method for calculating
gauge theory scattering amplitudes. This method, inspired by
Witten's twistor string \cite{Witten:2003nn} formulation of
$\mathcal{N}=4$ super-Yang--Mills (SYM), often produces very simple
expressions. While it is expected to be generally valid, to date
it has been successfully used at tree and one-loop
level \cite{Brandhuber:2004yw}. This expansion can in fact be
formally derived from the light-cone gauge fixed Yang-Mills action
\cite{Gorsky:2005sf, Mansfield:2005yd, Ettle:2006bw} and it can be
shown to be equivalent to the Feynman diagram expansion of the
twistor space Yang-Mills action \cite{Boels:2007qn}.

In closely related developments, by studying the analytic properties
of amplitudes as functions of complex external momenta, Britto et al
(BCFW) found recursion relations which generate all tree-level
amplitudes \cite{Britto:2004ap,Britto:2005fq}. The supersymmetric
generalizations of these relations
were solved explicitly \cite{Drummond:2008cr} giving relatively
compact expressions for all tree-level superamplitudes. By related
analytic methods in which all external lines are taken to be
complex, the MHV vertex expansion, \cite{Risager:2005vk,
Elvang:2008vz} was directly reproduced, thus showing that the BCFW
recursion relations and the MHV vertex expansion are equivalent at
tree level.


The  ${\cal N}=4$ SYM tree-level amplitudes constructed in
\cite{Drummond:2008cr} possess remarkable, hidden, conformal
symmetries \cite{Drummond:2008cr, Brandhuber:2008pf}. Hints of these
dual conformal symmetries were first seen in \cite{Drummond:2006rz},
and extended to dual superconformal symmetry in
\cite{Drummond:2008vq}.
For tree-level scattering amplitudes
there is in fact an enhancement of superconformal symmetry to an
infinite-dimensional algebra called the Yangian
\cite{Drummond:2009fd}.

The variables which make the dual conformal symmetry most apparent
are the ``momentum twistors'' introduced by Hodges
\cite{Hodges:2009hk}. These momentum twistors, which are
algebraically related to the null momenta of the amplitude and solve
the overall momentum constraint, are the spinors of the dual
conformal group.
Perhaps the most elegant formulation of the Yangian
invariants of  ${\cal N}=4$ SYM is the Grassmannian. The original
Grassmannian in standard twistor space was introduced in
\cite{ArkaniHamed:2009dn} where it was shown that a contour integral
over the Grassmannian $G(k,n)$ produced the N$^{k-2}$MHV
superamplitudes. An equivalent form in momentum twistor space was
soon found \cite{Mason:2009qx} which made the dual superconformal
symmetry manifest. That the Grassmannian has the full Yangian
symmetry was directly shown in \cite{Drummond:2010qh}. For
appropriate choices of the contour, the Grassmannian generates more
than tree-level amplitudes, indeed it was argued
\cite{Korchemsky:2010ut, Drummond:2010uq} that it generates all
Yangian invariants.

Beyond tree-level, IR divergences in amplitudes can no longer be
avoided and care must be taken in either regulating these
divergences or in choosing to study objects which are well defined
in spite of the bad IR behavior.
A set of such objects are the integrands of the loop integrals.
While in general such objects are ambiguous, as explained by
Arkani-Hamed et al \cite{ArkaniHamed:2010kv} (ABCCT), they can be
canonically defined in the planar limit. ABCCT further introduced a
recursive method (see also \cite{Boels:2010nw} for related work),
analogous to the BCFW recursion relations, for calculating the
all-loop integrand starting from tree amplitudes. These recursion
relations, in addition to providing an efficient method for
calculating the integrands, make their Yangian invariance manifest.

As was shown by Bullimore et al (BMS) \cite{Bullimore:2010pj}, the
MHV vertex expansion can also be usefully recast in momentum twistor
space, making the dual superconformal symmetry manifest. In this
formulation the ``propagators'' are dual superconformal invariants
while the vertices are simply unity. Using this formalism BMS gave
an algorithm for calculating any tree-level amplitude and any loop
integrand. As mentioned by BMS the expressions for the tree
amplitudes are very similar to those found by solving the BCFW
recursion relations and one might expect this to continue to be true
for the loop integrand. In this work we confirm that expectation.

We start in Sec. (\ref{sec:tree}), after a brief review of the momentum
space recursion relations and MHV vertex expansion in
Sec. (\ref{sec:review}), by showing that the tree amplitudes following from
the momentum twistor MHV expansion satisfy the BCFW recursion
relations. While this is in essence already known, we find that it
is a very useful warm-up for the loop integrand calculation as many
details are similar. In particular, while the MHV expansion
naturally produces sums over rooted tree diagrams, we find it useful
to recast it as sums over non-crossing partitions (which are well
known to be equivalent e.g. \cite{Stanley}). Turning to the loop
integrands in Sec. (\ref{sec:loop}) we explicitly study two examples -
one-loop MHV and one-loop NMHV - and show that the MHV expansions
satisfy the ABCCT recursion relations. We then introduce a rewriting
of the sum over diagrams for the MHV expansion for all legs and all
loops as a sum over a generalized class of non-crossing partitions,
and show that it satisfies the ABCCT recursion relations. The use of
non-crossing partitions enables us to write down the all-loop
integrands in an explicit and concise way, and it makes the full
classification of (graphs dual to) planar MHV diagrams clearer. As a
side result we find a recursive relation for the number of elements in a
certain class of non-crossing partitions, or equivalently, graphs dual to MHV
diagrams. In App.(\ref{app:diagrams}), we further explain the
one-to-one correspondence between non-crossing partitions and planar
MHV diagrams. By rewriting the expansion in
terms of dual graphs, the connections to the recently proposed dual
supersymmetric Wilson loops~\cite{Mason:2010yk, CaronHuot:2010ek}
become more transparent.

This result can be viewed in several ways. Firstly, while the MHV
vertex expansion is believed to be correct to all orders this has
never been proved beyond one-loop, this work can be seen as
providing evidence for the validity of the MHV expansion in showing
that it is equivalent to the recursion relations following from
analytic properties of the amplitudes. Secondly, it provides an explicit solution
to the recursion relations of ABCCT analogous to that found at
tree-level \cite{Drummond:2008cr}, An important feature of the
solution is that it is manifestly cyclic, thus it naturally unifies
different forms of all-loop integrands from ABCCT relations with
different shifts, and implies highly non-trivial relations between
them.

\section{A brief review of recursion relations and MHV vertex expansion in momentum-twistor space}
\label{sec:review}
In this section, we shall give a brief review of the recursion
relations and MHV vertex expansion in momentum-twistor space, for
all-loop integrands of scattering amplitudes in planar
$\mathcal{N}=4$ SYM. More details can be found
in~\cite{ArkaniHamed:2010kv, Bullimore:2010pj}.
\subsection{Momentum twistors for planar $\mathcal{N}=4$ SYM}
The $n$-particle superamplitude in planar $\mathcal{N}=4$ SYM,
$\mathcal{A}(\lambda_1,\tilde{\lambda}_1,\tilde{\eta}_1;\ldots;\lambda_n,\tilde{\lambda}_n,\tilde{\eta}_n)$,
depends on $n$ supermomenta, or equivalently, $n$ pairs of spinors
$(\lambda^{\alpha}_i,\tilde{\lambda}^{\dot{\alpha}}_i)$, and $n$
fermionic variable $\tilde{\eta}^I_i$, for $i=1,\ldots,n$.

Due to the color-ordering, one can define $n$ region supermomenta
$(x^{\alpha{\dot{\alpha}}}_i,\theta^{\alpha I}_i)$ by, \beqa
&x_i-x_{i+1}=\lambda_i\tilde{\lambda}_i,\nn
&\theta_i-\theta_{i+1}=\lambda_i\tilde{\eta}_i,\eeqa where
$(x_{n+1},\theta_{n+1})=(x_1,\theta_1)$, thus the supermomentum
conservation $\sum_i \lambda_i\tilde{\lambda}_i=0, \sum_i
\lambda_i\tilde{\eta}_i=0$, is automatically satisfied.

To make the dual superconformal symmetry of $\mathcal{N}=4$ SYM
manifest, one introduce $n$ momentum supertwistors,
\beqa \mathcal{Z}^A_i=\left( \begin{array}{ll} &\lambda^{\alpha}_i\\
&\mu^{\dot{\alpha}}_i\\
&\eta^{I}_i\\
\end{array} \right)=\left( \begin{array}{ll} &\lambda^{\alpha}_i\\
&x^{\alpha\dot{\alpha}}_i\lambda^{\alpha}_i\\
&\theta^{\alpha I}_i\lambda^{\alpha}_i\end{array} \right),\eeqa
which are defined projectively, for $i=1,\ldots,n$. The bosonic part
of the momentum supertwistor is denoted as $Z_i=(\lambda_i,\mu_i)$,
and henceforth we shall refer to $\mathcal{Z}_i$ simply as a
momentum twistor.

The inverse map is given by, \beqa
&x_i=\dfrac{\mu_i\lambda_{i-1}-\mu_{i-1}\lambda_i}{\langle
i-1,i\rangle},\nn
&\theta_i=\dfrac{\eta_i\lambda_{i-1}-\eta_{i-1}\lambda_i}{\langle
i-1,i\rangle},\eeqa where the 2-bracket is defined as $\langle
a,b\rangle=\epsilon_{\alpha\dot{\alpha}}\lambda^{\alpha}_a\tilde{\lambda}^{\dot{\alpha}}_b$.
Another set of useful relations express supermomenta in terms of
momentum twistors, \beqa &\tilde{\lambda}_i=\dfrac{\mu_{i-1}\langle
i,i+1\rangle+\mu_i\langle i+1,i-1\rangle+\mu_{i+1}\langle
i-1,i\rangle}{\langle i-1,i\rangle\langle i,i+1\rangle}\nn
&\tilde{\eta}_i=\dfrac{\eta_{i-1}\langle i,i+1\rangle+\eta_i\langle
i+1,i-1\rangle+\eta_{i+1}\langle i-1,i\rangle}{\langle
i-1,i\rangle\langle i,i+1\rangle}.\eeqa

The superamplitude $\mathcal{A}(1,\ldots,n)$ can be written in
momentum-twistor space by pulling out the MHV tree amplitude
$\mathcal{A}^{(0)}_{\rm{MHV}}(1,\ldots,n)$, \beq
\mathcal{A}(1,\ldots,n)=\mathcal{A}^{(0)}_{\rm{MHV}} A(1,\ldots,n).
\eeq Henceforth we shall work with $A(1,\ldots,n)$, which is dual
superconformal invariant. The geometric picture of amplitudes in
momentum-twistor space is simple. The point $x_i$ is associated with
the line $(i\,i+1)$ in momentum-twistor space, which is defined as
the line passing through two (bosonic) points in the
momentum-twistor space $Z_i$ and $Z_{i+1}$. Since the lines
$(i-1\,i)$ and $(i\,i+1)$ intersect at $Z_i$, it is guaranteed that
$x_i$ and $x_{i-1}$ are null separated, or equivalently, the
mass-shell conditions $p_i^2=(x_i-x_{i+1})^2=0$ are automatically
satisfied.

To see this explicitly, one can define the bosonic dual conformal
invariants through the  4-bracket,\footnote{The invariant is totally
anti-symmetric for the four twistors and so it can be defined in
terms of two lines, such as $(a\,b)$ and $(c\,d)$. }  $\langle
a,b,c,d\rangle=\langle
a\,b|c\,d\rangle=\epsilon_{ABCD}Z_a^AZ_b^BZ_c^CZ_d^D$, which
vanishes if and only if the four points are coplanar, or
equivalently, two lines, say, $(a\,b)$ and $(c\,d)$, intersect. For
the case $(a,b,c,d)=(i-1,i,j-1,j)$, the 4-bracket is related to the
Lorentz invariant, \beq (x_i-x_j)^2=\frac{\langle
i-1,i,j-1,j\rangle}{\langle i-1,i\rangle\langle j-1,j\rangle},\eeq
and we have seen that $(x_i-x_j)^2=0$ if and only if the lines
$(i-1\,i)$ and $(j-1\,j)$ intersect.

Furthermore, one can define the basic dual superconformal invariant
using five momentum supertwistors,\beq
[a,b,c,d,e]=\frac{\delta^{0|4}(\eta_a\langle
b,c,d,e\rangle+cyclic)}{\langle a,b,c,d\rangle\langle
b,c,d,e\rangle\langle c,d,e,a\rangle\langle d,e,a,b\rangle\langle
e,a,b,c\rangle}. \eeq As was shown in \cite{Mason:2009qx},  for the
special case $(a,b,c,d,e)=(n,i-1,i,j-1,j)$, it is given by the
invariant $R_{n;i,j}$, which appears in the NMHV amplitude
\cite{Drummond:2008cr}, \beq\label{NMHV1}
A^{(0)}_{\rm{NMHV}}=\sum_{1<i\prec j<n}R_{n;i,j}, \eeq where $i\prec
j$ means $i<j-1$. However, $[a,b,c,d,e]$ is defined for five general
momentum twistors, and we have seen that it has a simple pole when
four of the five twistors become coplanar. Also it is obvious that
the invariant is anti-symmetric for the five twistors, thus it
vanishes whenever two of them are identified projectively. As we
will see below, the invariant $[a,b,c,d,e]$ is the basic building
block of planar amplitudes in momentum-twistor space.

\subsection{BCFW recursion relations and loop generalizations in momentum-twistor space}

The BCFW recursion relations for tree-level superamplitudes in
$\mathcal{N}=4$ SYM have been reformulated in the momentum-twistor
space~\cite{ArkaniHamed:2010kv}. The BCFW deformation has a simple
form in momentum-twistor space, for which one needs to choose a
special leg, say, $\mathcal{Z}_1$, to shift,\footnote{Our choice of
deformation differs from that
in~\cite{ArkaniHamed:2010kv}.}\beq\label{BCFWdeform1}
\mathcal{Z}_1\rightarrow
\mathcal{Z}_{\hat{1}}=\mathcal{Z}_1+z\mathcal{Z}_2.\eeq Then the
color-ordered tree amplitude $M(1,\ldots,n)=A^{(0)}(1,\ldots,n)$ has
poles $z_j$ where $\langle n,\hat{1},j-1,j\rangle=0$, for
$j=4,\ldots,n-1$. The residue at the pole $z_j$ is given by the
``inhomogeneous'' term, $\int d^{0|4}\eta_{I_j}
M(\hat{1}_j,\ldots,j-1,I_j)\frac{1}{P^2_{I_j}} M(I_j,j,\ldots,n)$
where $\hat{1}_j$ and the internal leg $I_j$ is evaluated at the
pole, and it simplifies significantly in momentum-twistor
space~\cite{ArkaniHamed:2010kv}. In addition, there is a pole at
infinity, $\mathcal{Z}_{\hat{1}}\rightarrow \mathcal{Z}_2$
projectively when $z\rightarrow \infty$, and its residue is
non-vanishing.
\footnote{As pointed out in \cite{ArkaniHamed:2010kv}
there are also possible poles at $\langle Z_1(z)IZ_2\rangle$, with
$I$ the infinity tensor.  However
these poles would likely violate dual conformal symmetry
and so we do not expect them to contribute. In a theory without
dual conformal symmetry it is possible they would need to be included.}
 This corresponds to the ``homogeneous'' term with a
$3$-point anti-MHV amplitude attached to $M_{n-1}$, and in
momentum-twistor space, it is simply given by $M(2,\ldots,n)$.
Adding all pieces together, the recursion relations for tree-level
$n$-point N${}^k$MHV amplitudes
$M_{n,k}(1,\ldots,n)=A^{(0)}_{\rm{N}^k\rm{MHV}}(1,\ldots,n)$ can be
written as, \beqa \label{treerecursion}
M_{n,k}(1,\ldots,n)&=&M_{n-1,k}(2,\ldots,n)\\
& &+\sum_{j,k'}~ [j-1,j,n,1,2] ~M_{j,k'}(\hat{1}_j,\ldots,j-1,I_j)
M_{n+2-j,k-1-k'}(I_j,j,\ldots,n)~. \nonumber \eeqa The summation
ranges are $j=4,\ldots, n-1, k'=0,\ldots,k-1$ where terms with
$k'>j-4$ or $k-1-k'>n-j-2$ obviously vanish, and the deformations
are given by \beqa \mathcal{Z}_{\hat{1}_j}=\langle 2,j-1,j,n\rangle
\mathcal{Z}_1 +\langle j-1,j,n,1\rangle\mathcal{Z}_2
=(1\,2)\cap(j-1\,j\,n),\nn \mathcal{Z}_{I_j}= \langle
j,n,1,2\rangle\mathcal{Z}_{j-1}+ \langle n,1,2,j-1\rangle
\mathcal{Z}_j =(j-1\,j)\cap(n\,1\,2). \eeqa Note in the second
equalities we have adopted the geometric interpretation of the
deformations~\cite{ArkaniHamed:2010kv}, where $(i\,j)\cap(k\,l\,m)$
denotes the intersect of line $(i\,j)$ with plane $(k\,l\,m)$.


Loop amplitudes in $\mathcal{N}=4$ SYM suffer from IR divergences.
However, the integrands, which are not only IR-finite but simply
rational functions, can be unambiguously defined in the planar
sector and thus provide a well defined set of invariants which can
be calculated. In order to do just that, a generalized set of
BCFW-like recursion relations have been proposed for integrands for
all-loop amplitudes in planar $\mathcal{N}=4$
SYM~\cite{ArkaniHamed:2010kv}. To specify the integrands of $l$-loop
amplitudes, in addition to $n$ external momentum twistors, one also
needs $l$ pairs of momentum twistors $(A_m,B_m)$, for
$m=1,\ldots,l$, associated with $l$ loop momenta, upon which
integrations are performed. For the $n$-point $l$-loop N${}^k$MHV
superamplitude,
$A_{n,k,l}(1,\ldots,n)=A^{(l)}_{\rm{N}^{k}\rm{MHV}}(1,\ldots,n)$,
the integrand $M_{n,k,l}(1,\ldots,n)$ can be defined by the
generalized recursion relations~\cite{ArkaniHamed:2010kv}. The
formula for $l$-loop integrand $M_{n,k,l}$ is similar to the
tree-level one, plus a ``source term'' which comes from the poles
$\langle n,\hat{1},A,B\rangle=0$ for
$(A,B)=(A_1,B_1),\ldots,(A_l,B_l)$, \beqa\label{looprecursion}
M_{n,k,l}(1,\ldots,n;\{A,B\}_l)&=&M_{n-1,k,l}(2,\ldots,n;\{A,B\}_l)+\frac{1}{l!}\sum_{\sigma_l}\sum_{j,k'}~[n,1,2,j-1,j]~M^L~M^R\nn
&&+\frac{1}{l}\sum^l_{l_0=1}\int
_{\eta_{l_0}}\int_{GL(2)_{l_0}}[n,1,2,A_{l_0},B_{l_0}]M^S,
\eeqa
with
\beqa
& &M^L=M_{j,k',l'}(\hat{1}_j,\ldots,j-1,I_j;\{A,B\}_L)\nn
& & M^R=M_{n+2-j,k-1-k',l-l'}(I_j,j,\ldots,n;\{A,B\}_R)\nn
 && M^S=M'_{n+2,k+1,l-1}(\hat{1}_{A_{l_0}B_{l_0}},\ldots,n,A_{l_0},\hat{B}_{l_0};\{A,B\}_{\{l\}/l_0})
\eeqa
where in the second term on the r.h.s. of Eq.(\ref{looprecursion})
 we sum over all ways of distributing
$\{A,B\}_l$ into $\{A,B\}_L$ with $l'$ pairs of loop twistors and
$\{A,B\}_R$ with $l-l'$ pairs for $l'=0,\ldots,l$, and introduce an
$1/l!$ factor to compensate the overcounting; in the third term we
sum over $l_0$ and introduce an $1/l$ factor. The various
deformations are given by\beqa\label{BCFWdeform}
&\hat{1}_{j}=(1\,2)\cap(n\,j-1\,j),\, &I_j=(j-1\,j)\cap(n\,1\,2),\nn
&\hat{1}_{AB}=(1\,2)\cap(n\,A\,B),\,
&\hat{B}=(A\,B)\cap(n\,1\,2),\eeqa with $(A,B)=(A_{l_0},B_{l_0})$
for $l_0=1,\ldots,l$.

The source term, as given by the last line of
Eq.~(\ref{looprecursion}) is defined as the forward limit of
integrands at $l-1$ loop order. Given the prefactor and
$M_{n+2,k+1,l-1}$, the fermionic integration $\int_{\eta}=\int
d^{0|4}\eta_A d^{0|4}\eta_B$ removes two of the total $k+2$
fermionic delta functions to produce a result in the N$^k$MHV
sector. The ``$GL(2)$'' integration $\int_{GL(2)}$ is a contour
integral over general $GL(2)$ transformations $g$, which bring an
arbitrary $2$-vector $(A=\mathcal{Z}_A,B=\mathcal{Z}_B)$ to general
$2$-vectors $(A'=g_{11}A+g_{12}B,B'=g_{21}A+g_{22}B)$, and the
residue is given at the pole $A'\propto B'\propto \hat{B}$, or
geometrically when both $A$ and $B$ lie on the plane $(n,1,2)$. As we
shall explain in details shortly, the integrations in the source
term imply that the integrand only depends on the lines
$x_{(AB)_1},\ldots,x_{(AB)_l}$, or equivalently on the loop momenta,
$p_m=x_{(AB)_m}-x_{(AB)_{m+1}}$, for $m=1,\ldots,l$. Thus the
momentum-twistor space amplitudes are given by the usual loop
integrations, \beq A_{n,k,l}(1,\ldots,n)=\prod^l_{m=1}\int d^4
x_{(AB)_m}M_{n,k,l}(1,\ldots,n;\{x_{(AB)_1},\ldots,x_{(AB)_l}\}).\eeq

\subsection{MHV vertex expansion in momentum-twistor space}

Here we review the momentum-twistor space MHV vertex expansion for
$\mathcal{N}=4$ SYM as formulated in \cite{Bullimore:2010pj} and to
where we point the reader for further details. The $n$-point
N${}^k$MHV tree-level superamplitude $M_{n,k}$ is given by the sum
of all tree-level MHV diagrams with $k$ propagators and $k+1$ MHV
vertices. Each diagram is given as a product of factors from
vertices and propagators. Each vertex is given by unity, and for
each propagator separating region momenta $x_i,x_j$, one assigns a
factor $[\ast,\widehat{i-1},i,\widehat{j-1},j]$, where $\ast$ is an
arbitrary reference momentum twistor, and the possible deformation
$\widehat{i-1}$ is defined as (and similarly for $\widehat{j-1}$),
\beqa \widehat{i-1}=\left\{
\begin{array}{ll}
i-1, & \textit{if $i-1$ is attached to the propagator,}\\
\multirow{2}{*}{\mbox{$(i-1\,i)\cap(\ast\,k-1\,k)$},} &
\textit{otherwise, where $(k-1 k)$ is associated}\\ & \textit{~~with
the preceding propagator},
\end{array} \right.
\eeqa where since region momenta are ordered increasingly,
``preceding" means on the $i-1$ side. In~\cite{Bullimore:2010pj}, it
has been shown that by choosing
$\mathcal{Z}_{\ast}=(0,\iota^{\dot{\alpha}},0)$, the above rules
reproduce the usual momentum-space MHV diagrams with reference
spinor $\iota$.

Beyond tree level, in~\cite{Bullimore:2010pj}, the MHV vertex
expansion in the planar sector is conjectured to calculate the loop
integrands. The integrands defined by the recursion relations,
$M_{n,k,l}$, depend on external twistors and loop {\it momenta}, but the
integrands defined from the MHV vertex expansion,
$M'_{n,k,l}(1,\ldots,n; \{A_1,B_1,\ldots,A_l,B_l\})$, are dual
superconformal invariants, which generally depend on $n$ external
twistors and $l$ pairs of loop {\it twistors}. By fully integrating over
$2l$ loop momentum super-twistors, one obtains the loop amplitudes,
\beq\label{integral} A_{n,k,l}(1,\ldots,n)=\prod^l_{m=1}\int
d^{3|4}\mathcal{Z}_{A_m}d^{3|4}\mathcal{Z}_{B_m}M'_{n,k,l}(1,\ldots,n;\{A_1,B_1,\ldots,A_l,B_l\}).\eeq

Before reviewing the loop-level MHV vertex expansion we first explain
the relation between $M_{n,k,l}$ and $M'_{n,k,l}$. The integration
measure of a pair of loop twistors can be split as \beq
d^{3|4}\mathcal{Z}_Ad^{3|4}\mathcal{Z}_B=\langle\lambda_A
d\lambda_A\rangle\langle\lambda_B
d\lambda_B\rangle\langle\lambda_A\lambda_B\rangle^2
d^4x_{AB}d^{0|4}\eta_Ad^{0|4}\eta_B,\eeq where, in addition to the
fermionic integration measure, $d^4x_{AB}$ is the usual loop
integral measure, interpreted as a measure on the choice of lines
$x_{AB}$ through a pair of bosonic momentum twistors $Z_A,Z_B$, and
the measure of spinors is on the positions of $Z_A,Z_B$ on the line
$x_{AB}$. Now an equivalent way to integrate over positions on the
line is to integrate over all $GL(2)$ transformations, thus we can
write the integration over any pair of loop twistors $A,B$, as the
loop momentum integration, the fermionic integration, and the
$GL(2)$ integration, \beq\label{integration}\int
d^{3|4}\mathcal{Z}_Ad^{3|4}\mathcal{Z}_B M'_{n,k,l}(A,B)=\int
d^4x_{AB}\int d^{0|4}\eta_Ad^{0|4}\eta_B \int\langle g_1
dg_1\rangle\langle g_2dg_2\rangle\langle g_1g_2\rangle^2
M'_{n,k,l}(A',B')\, \eeq where the two $2$-vectors
$g_1=(g_{11},g_{12})$ and $g_2=(g_{21},g_{22})$ represent a $GL(2)$
transformation, $g$, that brings $(A,B)$ to $(A',B')=(A
,B)g^T=(g_{11} A+g_{12 }B,g_{21}A+g_{22} B)$.

Therefore, the relation between the two versions of integrands is,
\beq
M_{n,k,l}=\prod^l_{m=1}\int_{\eta_m}\int_{GL(2)_m}M'_{n,k,l},\eeq
where the fermionic and $GL(2)$ integrations for $A_m,B_m$, given by
Eq.~(\ref{integration}), have been denoted as $\int_{GL(2)_m}$ and
$\int_{\eta_m}$, respectively. For tree-level amplitudes
$M'_{n,k,0}=M_{n,k,0}=M_{n,k}$. At loop level, the complete result
from MHV diagrams in momentum-twistor space is supposed to be
independent of the positions of $Z_A,Z_B$ on the line $x_{AB}$,
which makes any $GL(2)$ integration trivial and yield a
constant~\cite{Bullimore:2010pj}. However, it is non-trivial to see the
independence of the result on the $GL(2)$ part of the integration term by term
in the momentum twistor formulation.
\footnote{We thank M. Bullimore for pointing out that this
independence is manifest in the standard momentum
space MHV vertex expansion.}

Now we briefly review the MHV vertex expansion for the loop
integrands $M'_{n,k,l}$ in momentum-twistor
space~\cite{Bullimore:2010pj}. The integrand $M'_{n,k,l}$ is given
by the sum of all $l$-loop planar MHV diagrams with $2l$ loop
twistors and cyclically ordered $n$ external twistors. Each MHV
diagram is given by the product of factors coming from $k+2l$
propagators and some vertices. Each vertex is again given by unity,
and each propagator is given by $[\ast,\,,\,,\,,\,]$ which depends
on a reference twistor and four other momentum twistors specified by
the following rules.

For a propagator separating regions of external momenta, $x_i$ and
$x_j$, the four twistors needed are
$\widehat{i-1},i,\widehat{j-1},j$, where the deformation is defined
as for tree amplitudes, except that the preceding region momentum
$(k-1\,k)$ can be a loop region momentum labeled by $A_m,B_m$. For a
propagator which separates $x_i$ from loop region $x_{(AB)_m}$, we
need $ \widehat{i-1},i,A_m,\widehat{B_m},$ with
$\widehat{B_m}=(A_m\,B_m)\cap(\ast\,k-1\,k)$ where $(k-1\,k)$ is the
preceding region momentum, which can be either an external or a loop
region. For a propagator separating $x_{(AB)_m}$ and
$x_{(AB)_{m'}}$, we need $A_m,\widehat{B_m},A_{m'},\widehat{B_{m'}}$
with similarly deformed loop twistors.

\section{Tree amplitudes}\label{sec:tree}
As a warm-up for loop-level integrands, in this section we will
prove that the MHV vertex expansion for tree amplitudes provides an
explicit solution to the BCFW recursion relations. After working out
some examples, we will rewrite the expansion for general tree
amplitudes and prove its validity.
\subsection{Examples}
\noindent
{\bf MHV}~~The MHV amplitude becomes trivial in
momentum-twistor space, $M_{n,0}=1$, which automatically satisfies
Eq.(\ref{treerecursion}) since there is no inhomogeneous terms.
\par
\begin{figure}
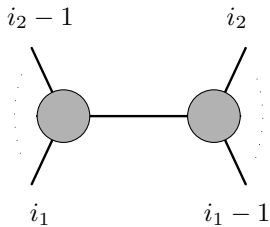
\centering
\includegraphicsbox[scale=1.0]{FigTreeNMHV.mps}
\caption{Vertex diagram for NMHV amplitude} \label{fig:FigTreeNMHV}
\end{figure}

\vspace{5mm}
\noindent
{\bf NMHV}~~For $k=1$, Fig. (\ref{fig:FigTreeNMHV}),
we have a summation over a pair of region momenta which label the
propagator, \beq\label{NMHV} M_{n,1}=\sum_{i_1\prec i_2\prec
i_1+n}[\ast,i_1-1,i_1,i_2-1,i_2],\eeq where no deformation is
needed, and instead of the standard summation range $i_1\leq i_2\leq
i_1+n$ we have used, for example, $i_1\prec i_2$ to denote
$i_1<i_2-1$ module $n$. Due to the total antisymmetry of the basic
invariant, terms with $i_1-i_2=0,\pm 1$ simply vanish. To compare
with Eq.~(\ref{treerecursion}), it is convenient to choose $\ast=n$,
then the summation range becomes $2\leq i_1\prec i_2\leq n-1$ and we
have,
\beq
M_{n,1}=\sum_{2\leq i_1\prec i_2\leq
n-1}[n,i_1-1,i_1,i_2-1,i_2].
\eeq This immediately follows from
Eq.~(\ref{treerecursion}) given that both subamplitudes must be MHV
amplitudes, that is to say, simply unity. This result is of course
nothing but the well known result Eq.~(\ref{NMHV1}). Note that the
MHV vertex expansion is independent of the choices of $\ast$, thus
Eq.~(\ref{NMHV}) provides a cyclically invariant explicit solution to
the BCFW recursion relations.
\par
\begin{figure}
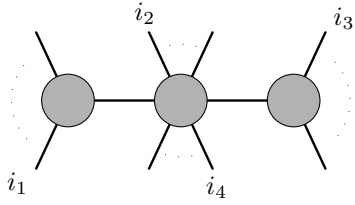
\centering
\includegraphicsbox[scale=1.0]{FigTreeNNMHV.mps}
\caption{Vertex diagram for NMHV amplitude} \label{fig:FigTreeNNMHV}
\end{figure}

\vspace{5mm}
\noindent {\bf N$^2$MHV}~~For $k=2$,  Fig.(\ref{fig:FigTreeNNMHV}),
the summation is over two pairs of region momenta, $i_1\prec i_2\leq
i_3\prec i_4\leq i_1+n$, \beq\label{NNMHV}
M_{n,2}=\sum_{i_1,i_2,i_3,i_4}[\ast,\widehat{i_1-1},i_1,i_2-1,i_2][\ast,\widehat{i_3-1},i_3,i_4-1,i_4],
\eeq where the summation range has been indicated before, and, from
the deformation rule, we know that when $i_1=i_4$ module $n$, a
deformation is needed,
$\widehat{i_1-1}=(i_1-1\,i_1)\cap(\ast\,i_3-1\,i_3)$, and when
$i_3=i_2$, $\widehat{i_3-1}=(i_3-1\,i_3)\cap(\ast\,i_1-1\,i_1)$
(these two boundary cases can not happen simultaneously since the
middle vertex must have at least three legs).

Again one chooses $\ast=n$, then for $M_{n,2}(1,\ldots,n)$ the
summation range can be split into two cases, $2\leq i_1\prec i_2\leq
i_3\prec i_4\leq n-1$ and $2\leq i_2\leq i_3\prec i_4\leq i_1\leq
n-1$, and similarly for $M_{n-1,2}(2,\ldots,n)$ with both lowest
limits being $3$. This corresponds to summing over the two
inequivalent ``rootings''of the ``1'' leg, i.e. on the first and
middle vertex. Then the difference in recursion relations is given
by terms in the first range with $i_1=2$ and those in the second
range with $i_2=2$ \beqa \label{eq:NMHV_diff}
M_{n,2}(1,\ldots,n)-M_{n-1,2}(2,\ldots,n)&=&\sum_{4\leq i_2\leq
i_3\prec i_4\leq
n-1}[n,1,2,i_2-1,i_2][n,\widehat{i_3-1},i_3,i_4-1,i_4]\nn
&&\kern-30pt+\sum_{2\leq i_3\prec i_4\leq i_1\leq
n-1}[n,\widehat{i_1-1},i_1,1,2][n,\widehat{i_3-1},i_3,i_4-1,i_4].
\eeqa
Now from Eq.~(\ref{NMHV}), with the choice $\ast=n$, we have for the
NMHV subamplitude \beq
M_{n+2-i_2,1}(I_{i_2},i_2,...,n)=\sum_{i_2\leq i_3\prec i_4\leq
n-1}[n,\widehat{i_3-1},i_3,i_4-1,i_4] \eeq because the deformation
needed when $i_3=i_2$ is exactly $\widehat{i_2-1}=I_{i_2}$. Thus the
first sum in Eq. (\ref{eq:NMHV_diff}) simply gives $\sum_{4\leq
i_2\leq i_3\prec i_4\leq
n-1}[n,1,2,i_2-1,i_2]M_{i_2,0}(1_{i_2},\ldots,i_2-1,I_{i_2})M_{n+2-i_2,1}(I_{i_2},i_2,\ldots,n)$.

For the second sum, we note that due to orientation reversal
symmetry~\cite{Bullimore:2010pj} we have the identity, \beq
[n,\widehat{i_1-1},i_1,1,2][n,\widehat{i_3-1},i_3,i_4-1,i_4]=[n,i_1-1,i_1,1,2][n,\widehat{i_3-1},i_3,i_4-1,\widehat{i_4}]
\eeq where $\widehat{i_4}=(i_1-1\,i_1)\cap(n\,1\,2)$ if $i_4=i_1$
and it is simply $i_4$ otherwise. Thus the second sum is equal to
$\sum_{2\leq i_3\prec i_4\leq i_1\leq
n-1}[n,1,2,i_1-1,i_1][n,\widehat{i_3-1},i_3,i_4-1,\widehat{i_4}]=\sum_{2\leq
i_3\prec i_4\leq i_1\leq
n-1}[n,1,2,i_1-1,i_1]M_{i_1,1}(1_{i_1},\ldots,i_1-1,I_{i_1})M_{n+2-i_1,0}(I_{i_1},i_1,...,n)$,
where we have used $\widehat{i_3-1}=1_{i_1}$ if $i_3=2$ and
$\widehat{i_4}=I_{i_1}$ when $i_4=i_1$.
Therefore, we see that Eq.~(\ref{NNMHV})also satisfies the recursion
relations.

\subsection{General tree amplitudes}

Generally for N${}^k$MHV tree amplitudes, in addition to summing
over the distributions of $n$ legs into $2k$ ordered subsets labeled
by region momenta, $i_1\leq i_2\leq \ldots \leq i_{2k}\leq i_1+n$,
one also needs to sum over all types of MHV diagrams. To be
concrete, one needs to pick a root, then there are
$C_k=\frac{(2k)!}{k!(k+1)!}$ (the Catalan number) types of diagrams
to be summed over, namely, all rooted trees with $k$ edges. It is
well known that they are in one-to-one correspondence with all
non-crossing partitions of the sequence
$i_1,i_2,\ldots,i_{2k-1},i_{2k}$ into $k$ pairs, where each pair
simply labels two region separated by an edge of the rooted tree.
Here ``non-crossing'' means that for the sequence $a,b,c,d$, one can
have $\{(a,b),(c,d)\}, \{(a,d),(c,b)\}$ as valid partitions but not
$\{(a,c),(b,d)\}$.

\begin{figure}
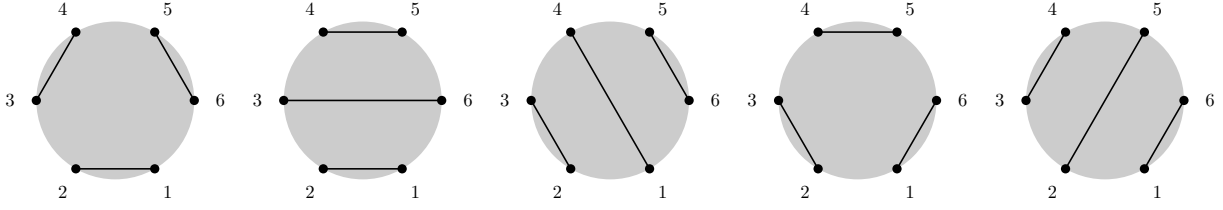
\centering
\includegraphicsbox[scale=0.7]{Tree_part_a.mps}~~
\includegraphicsbox[scale=0.7]{Tree_part_b.mps}~~
\includegraphicsbox[scale=0.7]{Tree_part_c.mps}~~
\includegraphicsbox[scale=0.7]{Tree_part_d.mps}~~
\includegraphicsbox[scale=0.7]{Tree_part_e.mps}~~
\caption{Elements of $J_3$ } \label{fig:TreePartitions}
\end{figure}

Each partition can be represented by a forest graph in which each
region is represented by a vertex and each pair by an edge. The
non-crossing partitions consists of the set of forests for which
edges only intersect at vertices. We denote the set of such
partitions, or equivalently rooted trees, as $J_k$.  For example,
$J_1=\{\{i_1,i_2\}\},J_2=\{\{i_1,i_2;i_3,i_4\},\{i_1,i_4;i_2,i_3\}\}$,
and
$J_3=\{\{i_1,i_2;i_3,i_4;i_5,i_6\},\{i_1,i_2;i_3,i_6;i_4,i_5\},\{i_1,i_4;i_2,i_3;i_5,i_6\},\{i_1,i_6;i_2,i_3;i_4,i_5\},\{i_1,i_6;i_2,i_5;i_3,i_4\}\}$.
As an example we show graphs corresponding to the elements of $J_3$
in Fig. (\ref{fig:TreePartitions}).
The relationship to geometric dual diagrams should be apparent and
we will make this more concrete later when we consider diagrams with
loops.

Given that the sequence $i_1,\ldots,i_{2k}$ labels all $2k$ ordered
region momenta, each pair in a non-crossing partition corresponds to
a propagator in a MHV diagram. Thus the tree amplitudes $M_{n,k}$
given by MHV vertex expansion can be written in an explicit form,
\beq\label{generaltree} M_{n,k}=\sum_{1\leq i_1\leq\ldots\leq
i_{2k}\leq n}\sum_{j^\alpha \in
J_k}\prod^{k}_{m=1}[\ast,\widehat{j^\alpha_{2m-1}-1},j^\alpha_{2m-1},\widehat{j^\alpha_{2m}-1},j^\alpha_{2m}],\eeq
where the first summation is over all distributions of legs,
\footnote{While this is the general summation range, inside each
type of the diagrams, a '$\leq$' will be replace by '$\prec$'
whenever two adjacent labels are separated by a propagator.} and the
second over all non-crossing partitions $j^\alpha$, where
$j^\alpha_{2m-1}$ and $j^\alpha_{2m}$ correspond to a region pair
$(m=1,\ldots,k)$ in the partition $j^\alpha$.
In addition, given a pair $j_{m+1}^\alpha=i_{l''}$,
$j^\alpha_m=i_l$, we have the definition:
$\widehat{j^\alpha_m-1}=(i_l-1\,i_l)\cap(\ast\,i_{l'}-1\,i_{l'})$ if
$i_l=i_{l-1}$, with  $(i_{l-1}, i_{l'})$ the pair in the partition
which precedes $(i_{l},i_{l''})$, and $\widehat{j^\alpha_m-1}=i_l-1$
otherwise.

As a final tree-level example, we write down the N${}^3$MHV
amplitude corresponding to Fig. (\ref{fig:TreePartitions}), denoting
$\prod^k_{m=1}[\ast,\widehat{j_{2m-1}-1},j_{2m-1},\widehat{j_{2m}-1},j_{2m}]$
by
$[\widehat{j_1},\widehat{j_2};\ldots;\widehat{j_{2k-1}},\widehat{j_{2k}}]$,
we have
 \beqa M_{n,3}& =&\sum_{i_1\leq\ldots\leq
i_6}[i_1,i_2;\widehat{i_3},i_4;\widehat{i_5},i_6]+[i_1,i_2;\widehat{i_3},\widehat{i_6};\widehat{i_4},i_5]+[i_1,\widehat{i_4};\widehat{i_2},i_3;\widehat{i_5},i_6]+[i_1,\widehat{i_6};\widehat{i_2},i_3;\widehat{i_4},i_5]
\nn & &\kern+25pt
+~[i_1,\widehat{i_6};\widehat{i_2},\widehat{i_5};\widehat{i_3},i_4].
\eeqa

\begin{figure}
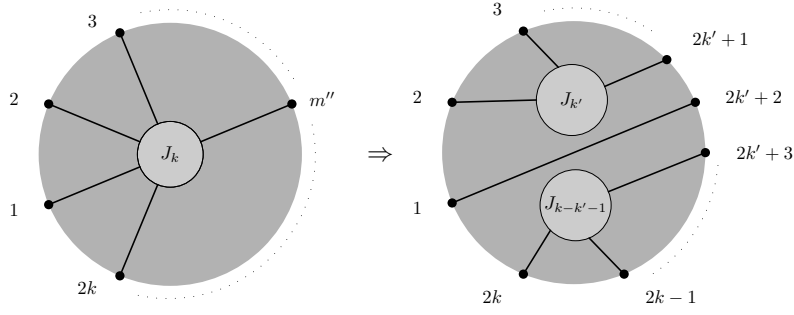
\centering
\includegraphicsbox[scale=0.7]{treefact_a.mps}~~$\Rightarrow$~~\includegraphicsbox[scale=0.7]{treefact_b.mps}
\caption{Factorization of non-crossing partition about the pair
$(i_1,i_{2m''})$.} \label{fig:facttree}
\end{figure}

Now it is straightforward to prove that Eq.~(\ref{generaltree})
provides an explicit solution to BCFW recursion relations,
Eq.~(\ref{treerecursion}). As always, it is convenient to choose
$\ast=n$, then taking the difference of $M_{n,k}(1,\ldots,n)$ and
$M_{n-1,k}(2,\ldots,n)$ we are left with all terms such that
$i_1=2$. The amplitude involves all non-crossing pairings of $i_1$
with all other regions, $i_2,i_4,\ldots,i_{2k}$. Given a choice of
pairing, say, $(i_1,i_4)$, subset of partitions in $J_k$ with this
choice consists of all products of  non-crossing partitions of $i_2,
i_3$ i.e. $J_1(i_2,i_3)$ and $i-5,\dots,i_{2k}$, i.e.
$J_{k-2}(i_5,\dots,i_{2k})$. More generally, $J_k$ can be split into
(with a slight abuse of notation), \beqa
J_k&=&\{i_1,i_2;J_{k-1}(i_3,\ldots,i_{2k})\}\cup\{i_1,i_4;J_2(i_2,i_3);J_{k-2}(i_5,\ldots,i_{2k})\}\cup\ldots\nn
& &~~~\dots\cup\{i_1,i_{2k};J_{k-1}(i_2,\ldots,i_{2k-1})\}. \eeqa
This is graphically shown in Fig.(\ref{fig:facttree}).
\footnote{The decomposition has a loop generalization, which is
shown in Fig. (\ref{fig:factloopF}) and here we only concern the
case with $l=0$.} In this way, the summation over $j^\alpha$ splits
into $k$ double summations, labeled by $k'=0,\ldots,k-1$. Each one
consists of a summation over $j'^\alpha\in
J_{k'}(i_2,\ldots,i_{2k'+1})$, a summation over $j''^\alpha\in
J_{k-k'-1}(i_{2k'+3},\ldots,i_{2k})$, and there is a prefactor
$[n,1,2,\widehat{i_{2k'+2}-1},i_{2k'+2}]$. The summation over
distributions of legs can be written as a summation over
$i_{2k'+2}$, a summation over $1<i_2\leq\ldots\leq i_{2k'+1}$, as
well as a summation over $i_{2k'+3}\leq \ldots\leq i_{2k}<n$. The
remaining factors also split into the product of two corresponding
pieces,
$\prod^{k'}_{m=1}[n,\widehat{j'^\alpha_{2m-1}-1},j'^\alpha_{2m-1},\widehat{j'^\alpha_{2m}-1},j'^\alpha_{2m}]$
and
$\prod^{k-1-k'}_{m=1}[n,\widehat{j'^\alpha_{2m-1}-1},j''^\alpha_{2m-1},\widehat{j''^\alpha_{2m}-1},j''^\alpha_{2m}]$.
\footnote{Note that in the case $k'=0~ (k-1)$, $J_{2k'}~
(J_{2(k-1-k')})$ is the empty set, and the first (second) piece is
given by unity.}

Therefore, the difference can be written as a sum of $k$ terms, each
being a product of two summations. From Eq.~(\ref{generaltree}), the
second sum is, \beqa && \sum_{i_{2k'+2}\leq i_{2k'+3}\leq \ldots\leq
i_{2k}<n}\sum_{j''^\alpha\in
J_{k-k'-1}}\prod^{k-1-k'}_{m=1}[n,\widehat{j''^\alpha_{2m-1}-1},j''^\alpha_{2m-1},\widehat{j''^\alpha_{2m}-1},j''^\alpha_{2m}]\nn
&
&\kern+160pt=M_{n+2-i_{2k'+2},k-1-k'}(\widehat{i_{2k'+2}-1},i_{2k'+2},\ldots,n),
\eeqa where one always needs the deformation
$\widehat{i_{2k'+2}-1}=(i_{2k'+2}-1\,i_{2k'+2})\cap(n,1,2)$, which
is exactly $I_{i_{2k'+2}}$ in Eq.~(\ref{BCFWdeform}). On the other
hand, due to the reversal symmetry, one can shift the deformation on
$i_{2k'+2}-1$ to $i_{2k'+1}$, which is
$\widehat{i_{2k'+1}}=(i_{2k'+2}-1\,i_{2k'+2})\cap(n,1,2)=I_{i_{2k'+2}}$,
thus the first summation is, \beqa &&\sum_{1<i_2\leq\ldots\leq
i_{2k'+1}\leq i_{2k'+2}}\sum_{j'^\alpha\in J_{k'}
}\prod^{k'}_{m=1}[n,\widehat{j'^\alpha_{2m-1}-1},j'^\alpha_{2m-1},\widehat{j'^\alpha_{2m}-1},j'^\alpha_{2m}]\nn
&&\kern+160pt=M_{i_{2k'+2},k'}(\widehat{1},\ldots,i_{2k'+2}-1,I_{i_{2k'+2}}),
\eeqa where $\widehat{1}=(1\,2)\cap(n\,i_{2k'+2}-1\,i_{2k'+2})$ is
$\hat{1}_{i_{2k'+2}}$ in Eq.~(\ref{BCFWdeform}). Therefore, we
obtain, \beqa
&&M_{n,k}-M_{n-1,k}=\sum_{k'=0,\ldots,k-1}\sum_{i_{2k'+2}}[n,1,2,i_{2k'+2}-1,i_{2k'+2}]\nn
\times
&&M_{i_{2k'+2},k'}(\hat{1}_{i_{2k'+2}},\ldots,i_{2k'+2}-1,I_{i_{2k'+2}})M_{n+2-i_{2k'+2},k-1-k'}(I_{i_{2k'+2}},i_{2k'+2},\ldots,n),\eeqa
which is nothing but the BCFW recursion relations,
Eq.~(\ref{treerecursion}).

Since the formula (\ref{generaltree}) is a rewriting of the
momentum-space MHV vertex expansion, it must be independent of the
reference twistor $\ast$. In fact, it is straightforward to check
that by choosing $\ast$ to be other external twistors, the formula
reproduces other forms of tree amplitudes, from BCFW recursion
relations with other shifts. As discussed in~\cite{Hodges:2009hk,
ArkaniHamed:2009dn}, it is highly non-trivial to prove the
equivalence of these different forms, which guarantees important
properties, such as cyclic invariance and absence of spurious poles,
of the tree amplitudes. It is significant that the formula obtained
here relates different BCFW forms of the tree amplitude, and it is
manifestly cyclically invariant.

\section{All-loop Integrands}\label{sec:loop}

Now we move to integrands of all-loop amplitudes in the planar
sector. The MHV vertex expansion becomes more involved at higher
loops, so we will first examine carefully the one-loop MHV and NMHV
integrands as solutions to the generalized recursion relations. Then
we propose and prove a general formula for the MHV vertex expansion
in terms of non-crossing partitions, as a straightforward
generalization of the tree-level case.

\subsection{Examples of Loop Integrands}

\subsubsection{One-loop MHV integrand}
\begin{figure}
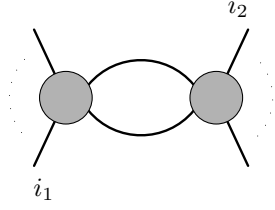
\centering
\includegraphicsbox[scale=1.0]{FigOneloopMHV.mps}
\caption{Vertex diagram for one-loop MHV amplitude}
\label{fig:OneloopMHV}
\end{figure}

The MHV one-loop integrand, $M'_{n,0,1}$, which only receives
contribution from the ``bubble" topology, is given by,
Fig.(\ref{fig:OneloopMHV}),
\beq\label{1loopMHV}
M'_{n,0,1}=\sum_{i_1<i_2}[\ast,i_1-1,i_1,A,B'][\ast,i_2-1,i_2,A,B''],\eeq
where $B'=(A\,B)\cap(\ast\,i_2-1\,i_2)$ and
$B''=(A\,B)\cap(\ast\,i_1-1\,i_1)$. We shall prove that
Eq.~(\ref{1loopMHV}), after doing fermionic and $GL(2)$
integrations, provides an explicit solution to
Eq.~(\ref{looprecursion}). Each term of $M'_{n,0,1}$ only depends on
the line $(A,B)$ and we note that $(A\,B)=(A\,B')$. Since
$B''=(A\,B')\cap(\ast\,i_1-1\,i_1)$, and the fermionic and $GL(2)$
integration measure in $M_{n,0,1}$ are invariant under the shift
$B\rightarrow B'$, we have that (using the relation between
integrands $M$ and $M'$)
 \beq
M_{n,0,1}=\sum_{ i_1<i_2}\int_{\eta_{A{\tilde
B}'}}\int_{GL(2)_{A{\tilde B}'}}[\ast,i_1-1,i_1,A,{\tilde
B}'][\ast,i_2-1,i_2,A,B'']~. \eeq Here it is important to clarify
the integration contour. The $GL(2)$ integration relating
$M'_{n,0,1}$ to $M_{n,0,1}$ is trivial. The $GL(2)_{A{\tilde B}'}$
integration above should be understood as a contour integration
forcing the {\it dummy variable} ${\tilde B}'$ to lie in the plane
defined by $(\ast\,i_2-1\,i_2)$ i.e. at the point $B'$.
 Now since after $GL(2)$ integration the result only depends on
$x_{AB}$ which is also $x_{A{\tilde B}'}$, we can shift it back, \beq
M_{n,0,1}=\sum_{
i_1<i_2}\int_{\eta}\int_{GL(2)}[\ast,i_1-1,i_1,A,B][\ast,i_2-1,i_2,A,B''].
\eeq Note that this is actually just a change of dummy variable
because after integration $M_{n,0,1}$ only depends on $B$ through
$x_{AB}$, which satisfies $x_{AB}=x_{AB'}$.

If we choose $\ast=n$, then the difference between $M_{n,0,1}$ and
$M_{n-1,0,1}$ is given by terms with $i_1=2$, \beq
M_{n,0,1}-M_{n-1,0,1}=\int_{\eta}\int_{GL(2)}[n,1,2,A,B]\sum_{
2<i_2<n}[n,i_2-1,i_2,A,\hat{B}],\eeq where we have used
$B''=\hat{B}$ for $i_1=2$ and $\ast=n$. The only contribution in
Eq.~(\ref{looprecursion}) for this case is the source term with
$M_{n+2,1,0}(1_{AB},\ldots,n,A,\hat{B})$. As we will show shortly,
in the NMHV tree amplitudes Eq.~(\ref{NMHV}) with $\ast=n$, only
terms which have at least one factor with both $A$ and $\hat{B}$ as
arguments survive the fermionic integration. We conclude that the
one-loop MHV integrand from the MHV vertex expansion satisfies the
recursion relations.

We can in fact go further in this case as it is straightforward to
explicitly perform the fermionic and $GL(2)$ integrations, \beqa
\label{1loopMHVexplicit}
M_{n,0,1}
&=&\sum_{1<i_1<i_2<n}\Big[ \langle
A\,B|(n\,i_1-1\,i_1)\cap(n\,i_2-1\,i_2)\rangle^2\nn &
&\kern-60pt\times \frac{1}{\langle A,B,i_1-1,i_1\rangle\langle
A,B,i_1-1,n\rangle\langle A,B,i_1,n\rangle\langle
A,B,i_2-1,i_2\rangle\langle A,B, i_2-1,n\rangle\langle A,B,
i_2,n\rangle}\Big], \eeqa where in the numerator $\langle
AB|(n\,i_1-1\,i_1)\cap(n\,i_2-1\,i_2)\rangle=\langle A,
n,i_1-1,i_1\rangle\langle B,n,i_2-1,i_2 \rangle-\langle
B,n,i_1-1,i_1 \rangle\langle A,n,i_2-1,i_2\rangle$, with
$(n\,i_1-1\,i_1)\cap(n\,i_2-1\,i_2)$  the intersecting line of the
two planes, and, as promised, the $GL(2)$ integration renders each
term only a function of line $x_{AB}$.

\subsubsection{A lemma for the source term}

Here we prove a lemma for the source term: in the MHV vertex
expansion of the integrand
$M_{n+2,k+1,l-1}(\hat{1}_{AB},\ldots,n,A,\hat{B})$, only those
terms, which have at least one factor with both loop twistors and no
factor with a single loop twistor, survive the fermionic
integration. This is crucial for relating MHV vertex expansion and
recursion relations, since no other terms should appear in the MHV
vertex expansion.

First, any factor with a single loop twistor, $A$ or $\hat{B}$,
vanishes by itself. For convenience, we choose $\ast=n$ for the MHV
vertex expansion of
$M_{n+2,k+1,l-1}(\hat{1}_{AB},\ldots,n,A,\hat{B})$, then invariants
with only $A$ vanish, $[n,x,y,n,A]=0$ for any $x,y$, and thus we
focus on the factors with only $\hat{B}$, i.e.
$[n,x,y,\hat{B},\hat{1}_{AB}]$.

Since $\hat{B}=(A\,B)\cap(n\,1\,2)$ and
$\hat{1}_{AB}=(1\,2)\cap(n\,A\,B)$, $n, \hat{B}$ and $\hat{1}_{AB}$
must lie on the same line in the momentum-twistor space, and it is
straightforward to write down the linear dependence, \footnote{The
easiest way to see it is the following: the sum of the first two
terms give $(n\,\hat{B})\cap(\hat{1}_{AB}\,x\,y)$ which is
projectively $\hat{1}_{AB}$ since the three are on a line, thus the
L.H.S. is proportional to $\hat{1}_{AB}$; the sum of the last two
terms give $(\hat{B}\,\hat{1}_{AB})\cap(n\,x\,y)$ which is
projectively $n$, thus the L.H.S. is proportional to $n$, and we
conclude it must vanish.} \beq \langle
x,y,\hat{B},\hat{1}_{AB}\rangle\mathcal{Z}_n+\langle
\hat{1}_{AB},n,x,y\rangle\mathcal{Z}_{\hat{B}}+\langle
n,x,y,\hat{B}\rangle\mathcal{Z}_{\hat{1}_{AB}}=0,\eeq for any $x,y$.
In the fermionic delta function of $[n,x,y,\hat{B},\hat{1}_{AB}]$,
the sum of the three terms with $\eta_n,\eta_{\hat{B}}$ and
$\eta_{\hat{1}_{AB}}$ vanishes since it is the fermionic part of the
above expression, then we have, \beq
[n,x,y,\hat{B},\hat{1}_{AB}]=\frac{\delta^{0|4}(\langle
y,\hat{B},\hat{1}_{AB},n\rangle\eta_x+\langle\hat{B},\hat{1}_{AB},n,x\rangle\eta_y)}{\langle
\hat{1}_{AB},n,x,y\rangle\langle n,x,y,\hat{B}\rangle\langle
x,y,\hat{B},\hat{1}_{AB}\rangle\langle
y,\hat{B},\hat{1}_{AB},n\rangle\langle
\hat{B},\hat{1}_{AB},n,x\rangle},\eeq where, due to the linear
dependence of $n,~\hat{B}$ and $\hat{1}_{AB}$, the numerator has
four zeros from e.g. $\langle y,\hat{B},\hat{1}_{AB},n\rangle^4$ or
$\langle\hat{B},\hat{1}_{AB},n,x\rangle^4$, and the denominator has
two zeros from the last two factors, so it vanishes identically.

In addition, for any term to survive the fermionic integration over
$\eta_A,\eta_B$, one needs four $\eta_A$ and four $\eta_B$, which
means the term must have at least two factors containing (possibly
deformed) loop twistors. Since there is a prefactor $[n,1,2,A,B]$,
this excludes any term in $M_{n+2,k+1,l-1}$ which has no factor with
any loop twistors. Thus we have seen, only terms which have
factor(s) with both $A$ and $\hat{B}$ and no factor with one of
them, $A$ or $\hat{B}$, survive the fermionic integration.

\subsubsection{One-loop NMHV integrand}

\begin{figure}
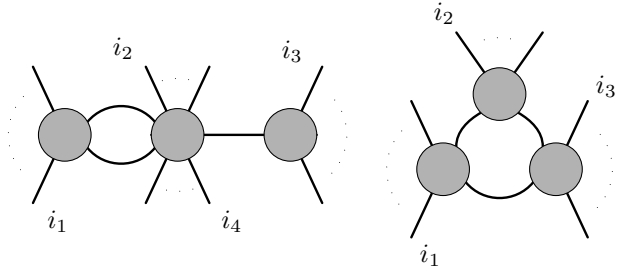
\centering
\includegraphicsbox[scale=1.0]{FigOneloopNMHV_a.mps}~~~
\includegraphicsbox[scale=1.0]{FigOneloopNMHV_b.mps}
\caption{Bubble $+$ leg and triangle diagrams for one-loop NMHV
amplitude} \label{fig:OneloopNMHV}
\end{figure}

The next simplest case is the one-loop NMHV integrand, $M'_{n,1,1}$.
The MHV vertex expansion has contributions from both ``triangle" and
``bubble $+$ leg" topologies see Fig.(\ref{fig:OneloopNMHV}), and
the result is,
 \beqa\label{1loopNMHV}
&&M'_{n,1,1}=\sum_{i_1<i_2<i_3}[\ast,i_1-1,i_1,A,B'][\ast,i_2-1,i_2,A,B''][\ast,i_3-1,i_3,A,B''']\nn
&&+\sum_{i_1<i_2\leq i_3\prec
i_4}[\ast,\widehat{i_1-1},i_1,A,B^*][\ast,i_2-1,i_2,A,B^{**}][\ast,\widehat{i_3-1},i_3,i_4-1,i_4],\eeqa
where in the first sum, $B'=(A\,B)\cap(\ast\,i_3-1\,i_3),
B''=(A\,B)\cap(\ast\,i_1-1\,i_1)$, and
$B'''=(A\,B)\cap(\ast\,i_2-1\,i_2)$; in the second sum, deformations
are needed, $\widehat{i_1-1}=(i_1-1\,i_1)\cap(\ast\,i_4-1\,i_4)$
when $i_1=i_4$ module $n$,
$\widehat{i_3-1}=(i_3-1\,i_3)\cap(\ast\,A\,B'')$ when $i_3=i_2$, and
we also have $B^*=(A\,B)\cap(\ast\,i_2-1\,i_2),
B^{**}=(A\,B)\cap(\ast\,\widehat{i_1-1}\,i_1)$.

Now we prove that Eq.~(\ref{1loopNMHV}) also satisfies the recursion
relations. Choosing $\ast=n$, and noting that in each term of the
summation we can use the same trick as in the MHV case to replace
$B',{B^\ast}$ or ${B^\ast}$ by $B$ without changing anything else,
then the difference of $M_{n,1,1}$ and $M_{n-1,1,1}$ is given by
$i_1=2$ terms in the first sum, and terms with $i_1=2$ or  $i_2=2$
or $ i_3=2$ or $i_4=2$ in the second sum, \beqa
&&M_{n,1,1}-M_{n-1,1,1}=\int_{\eta}\int_{GL(2)}\{ \sum_{2<i_2<i_3<
n}[n,1,2,A,B][n,i_2-1,i_2,A,B''][n,i_3-1,i_3,A,\widehat{B''}]\nn
&&+\sum_{2<i_2\leq i_3\prec
i_4<n}[n,1,2,A,B][n,i_2-1,i_2,A,B^{**}][n,\widehat{i_3-1},i_3,i_4-1,i_4]\nn
&&+\sum_{2\leq i_3\prec i_4\leq
i_1<n}[n,\widehat{i_1-1},i_1,A,B^*][n,1,2,A,B][n,\widehat{i_3-1},i_3,i_4-1,i_4]\nn
&&+\sum_{3<i_4\leq
i_1<i_2<n}[n,\widehat{i_1-1},i_1,A,B^*][n,i_2-1,i_2,A,B^{**}][n,1,2,i_4-1,i_4]\nn
&&+\sum_{2\leq i_1<i_2\leq
i_3<n}[n,\widehat{i_1-1},i_1,A,B^*][n,i_2-1,\widehat{i_2},A,B^{**}][n,i_3-1,i_3,1,2]\},\eeqa
where in the first line we have replaced
$B'''=(A\,B)\cap(n\,i_2-1\,i_2)$ by
$\widehat{B''}=(A\,B'')\cap(n\,i_2-1,i_2)$ since $(A\,B)=(A\,B'')$,
and in the last line we have used the reversal symmetry to shift the
deformation from $i_3-1$ to $i_2$.

Since in the first three lines, $B''=B^\ast=B^{\ast\ast}=\hat{B}$
and $\widehat{B''}=\hat{\hat{B}}_{i_2}$ which have been defined in
Eq.~(\ref{BCFWdeform}), from Eq.~(\ref{NNMHV}) we immediately
recognize that these terms appear in
$M_{n+2,2,0}(\hat{1}_{AB},\ldots,n,A,\hat{B})$, and all other terms
without $A$ or $\hat{B}$ simply vanish upon the fermionic
integration, thus these three lines combine to the source term,
$\int_{\eta}\int_{GL(2)}[n,1,2,A,B]
M_{n+2,2,0}(\hat{1}_{AB},\ldots,n,A,\hat{B})$. In the last two
lines, $B^\ast$ and $B^{\ast\ast}$ are the same as in
Eq.~(\ref{1loopMHV}), and $\widehat{i_1-1}=\hat{1}_{i_4},
\hat{1}_{i_3}$ when $i_1=2$ in the fourth and the last line,
respectively, also $\hat{i_2}=I_{i_3}$, thus, after the fermionic
and $GL(2)$ integration, they combine to the factorization term
$\sum_j M_{j,0,1}M_{n+2-j,0,0}+M_{j,0,0}M_{n+2-j,0,1}$. Therefore,
one-loop NMHV integrand from MHV vertex expansion also satisfies
recursion relations.

\subsection{All-loop integrands}

Here we propose an explicit formula for all-loop integrands from MHV
vertex expansion, \beq\label{eq:allloop}
M'_{n,k,l}=\frac{1}{l!}\sum_{j^{\alpha}\in J^{l}_{k}}\sum_{1\leq
i_1\leq\ldots\leq i_m\leq n}\prod_{e\in E(j^{\alpha})}(\pm)
[*,\widehat{v_1(e)-1},v_1(e),\widehat{v_2(e)-1},v_2(e)]. \eeq Here
$J^l_k$ is the set of all non-crossing partitions
$j^{\alpha}(X,I,E,F)$. Each partition $j^{\alpha}$ is a forest with
the following elements: $m$ cyclically ordered \footnote{We will
consider diagrams where all external points are distinct. However we
allow the sum over regions to include degenerate cases
$i_{l}=i_{l-1}$. Alternatively, one could allow the vertices to
coincide but restrict the sum to be over distinct $i_{l}<i_{l-1}$.}
external points corresponding to region momenta,
$X=\{i_1,\ldots,i_m\}$, where each point has one edge connected to
it; $l$ internal points corresponding to loop momenta,
$I=\{A_1,\ldots,A_l\}$, where multiple edges can attach to the same
point; \footnote{While the graphs can have any number of edges
coincident with an internal vertex for the amplitudes only those
partitions with at least two {\it distinct} edges attached to each
internal vertex will give a non-vanishing contribution. This is
essentially due to the fermionic integration and in part corresponds
to the absence of tadpole diagrams in the MHV expansion.} $k+2l$
non-crossing edges corresponding to propagators, $E$, where each
edge $e$ is associated with its two vertices, $v_j(e)\in X\cup I$
for $j=1,2$, and $f$ faces formed by internal points and edges
connecting them, $F$. By definition, $J^l_k$ includes all partitions
with all possible permutations of $l$ internal points,
$\{A_1,\ldots,A_l\}$. From Eq.~(\ref{integral}), the $l$ pairs of
loop twistors are dummy variables in the integrated amplitude, and
any permutation gives the same result, thus one needs an $1/l!$
factor to compensate the overcounting .

It is obvious that $f=0$ for $l\leq 2$, and for $l\geq 3$, there can
be faces with the range $0\leq f\leq k+l-2$. If $f=0$, the range of
$m$ is $k+l+1\leq m\leq 2k+2l$, while for $f>0$, we have
$k+l-f+1\leq m\leq 2k+2l-2f+1$. The set of partitions with $l$
internal points and $k+2l$ edges has been denoted by $J_{k}^{l}$.
This set of diagrams is the union of sets of diagrams with $l$
internal points, $k+2l$ edges, $m$ external points,
$J^{l}_k=\cup_{m=2}^{2k+2l} J^{l}_{k;m}$. We can further decompose
the graphs by the number of faces, $f$,
$J^{l}_{k;m}=\cup_{f=0}^{k+l+1-m} J^{l}_{k;m,f}$. We define
$J^{l}_{k;m,f}$ so that all external points are distinct,
thus each external leg is directly connected to a single edge and as
mentioned $f\geq1$ only for $l\geq 3$.
These partitions are in one-to-one correspondence with the planar
MHV diagrams via their dual graphs as we explain in
App.(\ref{app:diagrams}) in slightly more detail. These dual graphs
have appeared in this context recently in \cite{Mason:2010yk,
Brandhuber:2010mi}.

Given any such partition, we sum over all possible distributions of
$n$ legs into $m$ ordered intervals, $1\leq i_1\leq\ldots\leq
i_m\leq n$, where it is obvious that some degenerate cases drop out.
In the product of edges, a minus sign is needed when one of the two
vertices are one of the internal points, because we define the
internal points to be $A$, $v_j(e)=A_{l'}$, and then
$v_j(e)-1=B_{l'}$. In addition, $\widehat{v_j(e)-1}=v_j(e)-1$ when
$v_j(e)=i_{m'}\in X$ for some $m'$ and $i_{m'}\neq i_{m'-1}$; a
shift is needed,
$\widehat{v_j(e)-1}=(v_j(e)-1\,v_j(e))\cap(*\,v-1\,v)$ when either
$v_j(e)=i_{m'}=i_{m'-1}$ or $v_j(e)\in I$, where $v$ is the other
vertex of the preceding edge $e'$ which shares $v_j(e)$ with $e$.

\subsubsection{Examples of Partitions}
\begin{figure}
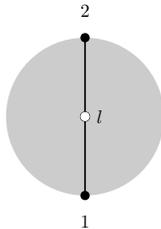
\centering
\includegraphicsbox[scale=0.7]{One_loop_MHV_part.mps}
\caption{One-loop MHV partition.}
\label{fig:One_loop_MHV_Partitions}
\end{figure}

\begin{figure}
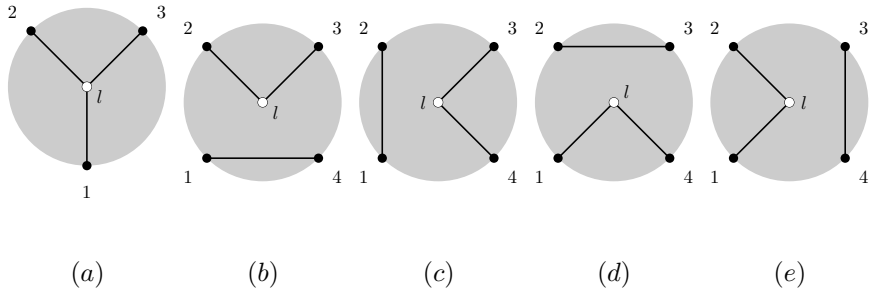
\centering
\begin{eqnarray}
\begin{array}{ccccc}
\includegraphicsbox[scale=0.7]{One_loop_NMHV_part_a.mps} &
\includegraphicsbox[scale=0.7]{One_loop_NMHV_part_b.mps} &
\includegraphicsbox[scale=0.7]{One_loop_NMHV_part_c.mps} &
\includegraphicsbox[scale=0.7]{One_loop_NMHV_part_d.mps} &
\includegraphicsbox[scale=0.7]{One_loop_NMHV_part_e.mps}\nn \\
 (a)&  (b)&  (c) & (d)  & (e)
\end{array}
\end{eqnarray}
\caption{One-loop NMHV partitions}
\label{fig:One_loop_NMHV_Partitions}
\end{figure}

\begin{figure}
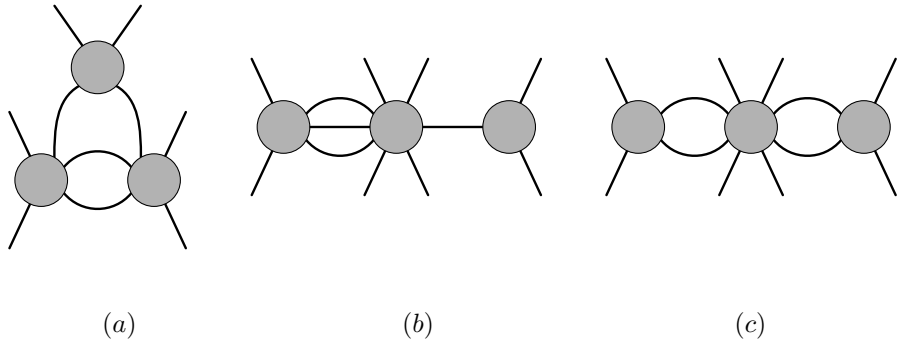
\centering
\begin{eqnarray}
\begin{array}{ccc}
\includegraphicsbox[scale=1.0]{FigTwoloopMHV1.mps}~~~~~&
\includegraphicsbox[scale=1.0]{FigTwoloopMHV2.mps}~~~~~ &
\includegraphicsbox[scale=1.0]{FigTwoloopMHV3.mps}\nn \\
 (a)&  (b)&  (c) \nn
\end{array}
\end{eqnarray}
\caption{Two-loop MHV diagrams} \label{fig:Two_loop_MHV_diagrams}
\end{figure}

\begin{figure}
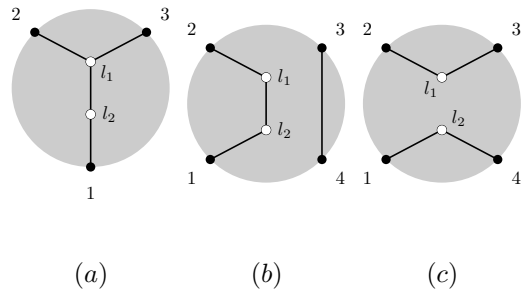
\centering
\begin{eqnarray}
\begin{array}{ccc}
\includegraphicsbox[scale=0.7]{Two_loop_MHV_part_a.mps} &
\includegraphicsbox[scale=0.7]{Two_loop_MHV_part_b.mps} &
\includegraphicsbox[scale=0.7]{Two_loop_MHV_part_c.mps} \nn\\
 (a)&  (b)&  (c)
\end{array}
\end{eqnarray}
\caption{Two-loop MHV partitions}
\label{fig:Two_loop_MHV_Partitions}
\end{figure}

We use a few examples to illustrate Eq.~(\ref{eq:allloop}). At tree
level, the partitions  have $k$ edges, no internal points, and so
must have $2k$ external points, i.e. $J^0_k=J^0_k|_{m=2k}$ and it is
simply given by the $J_{k}$ which occurred at tree level, where each element can
be represented by a diagram
as in Fig.(\ref{fig:TreePartitions}), and Eq.~(\ref{eq:allloop}) reduces to Eq.~(\ref{generaltree}) in this case. 
%
%
At loop level we must include internal points in our partition
diagrams. The simplest case, one-loop MHV has only one element,
$J^1_0=\{j\}$ with $X(j)=\{i_1,i_2\}, I(j)=\{i_l\},
E(j)=\{i_1,i_l;i_2,i_l\}$, which corresponds to the diagram in
\ref{fig:One_loop_MHV_Partitions}, and we have exactly
Eq.~(\ref{1loopMHV}).
For one-loop NMHV we have $J^1_1=\{j^a,j^b,j^c,j^d,j^e\}$, where
$j^a$ has $m=3$ (see \ref{fig:One_loop_NMHV_Partitions} (a)) and
$j^b$ to $j^e$ have $m=4$ (see \ref{fig:One_loop_NMHV_Partitions}
(b),(c),(d),(e)), and the integrand is given by the sum of five
contributions, Eq.~(\ref{1loopNMHV}). This can be straightforwardly
continued to higher $k$.
Similarly one can continue to higher-loop; for the two-loop MHV
diagram the set $J^2_0$ has elements shown in
\ref{fig:Two_loop_MHV_Partitions} (plus permutations) with $m=3,4$
which correspond to the MHV diagrams Fig.
(\ref{fig:Two_loop_MHV_diagrams}). The diagrams $(a)$ and $(c)$ are
those listed in~\cite{Bullimore:2010pj}, but diagram $(b)$ is also a
valid partition, though one which of course vanishes
under the $\eta$ integration.
%

\subsubsection{Proof}

Now we proceed to prove inductively that Eq.~(\ref{eq:allloop})
satisfies the generalized recursions relations. This provides strong
evidence that it is indeed a correct expression for the all-loop
integrand.

By assumption Eq.~(\ref{eq:allloop}) is independent of the
$\ast$, and so one again chooses $\ast=n$ for convenience. Taking
the difference $\int_{\{l\}}
(M'_{n,k,l}(1,\ldots,n;\{A,B\}_{\{l\}})-M_{n-1,k,l}(2,\ldots,n;\{A,B\}_{\{l\}})$),
where the $l$-fold fermionic and $GL(2)$ integrations have been
denoted implicitly, only terms with $i_1=2$ survive. Thus we pick
out the edge connecting $i_1$ to another vertex $v$.  There are two
possibilities, $v\in X$ and $v\in I$, which we now discuss in turn.

\begin{figure}
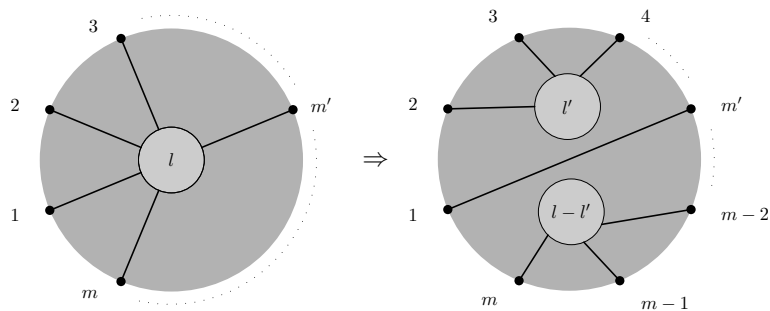
\centering
\includegraphicsbox[scale=0.7]{Loopfact_a.mps}~~$\Rightarrow$~~\includegraphicsbox[scale=0.7]{Loopfact_b.mps}
\caption{Factorization of loop diagram of type $F$.}
\label{fig:factloopF}
\end{figure}

For a partition $j^\alpha\in J^l_k$ with $m$ external legs and with
$v=i_{m'} ~(m'=2,\ldots,m)$, one can pull out a factor from the set
of edges $e_0=(i_1,i_{m'})$, corresponding to  the invariant
$[n,1,2,\widehat{i_{m'}-1},i_{m'}]$, for each $m'$, Fig.
(\ref{fig:factloopF}). This gives rise to the terms
 \beq F=\frac{1}{l!}\sum_{j^\alpha\in
J^l_k}\sum^m_{m'=2}\sum^{n-1}_{i_{m'}=4}\int_{\{l\}}\sum_{1<i_2\leq\ldots\leq
i_{m'-1}\leq i_{m'}}\sum_{i_{m'}\leq i_{m'+1}\leq \ldots\leq
i_m<n}\prod_{e\neq e_0}[e]~, \eeq where we denote the invariants as
$[e]=\pm[n,\widehat{v_1(e)-1},v_1(e),\widehat{v_2(e)-1},v_2(e)]$.

Now one can rewrite the summation over the partitions for the
remaining $k+2l-1$ factors as a summation over $J^l_{k-1}(m')$,
which is defined as the set of all partitions in $J^{l}_{k}$ with
$i_1,i_{m'}$ and the edge connecting them removed. By the definition
of $J^l_k$, each partition $j\in J^l_{k-1}(m')$ splits into two
sub-partitions $j_L$ and $j_R$, both of which are non-crossing
partitions. For a partition $j$ with $m$ external legs, the first
subpartition,  $j_L\in J^{l'}_{k';m'-2}$, has  $m'-2$  external legs
while $j_R\in J^{l-l'}_{k-1-k';m-m'}$ has $m-m'$ external legs. Note
that the total number of internal points is $l$, and the total
number of edges is $k'+2l'+(k-1-k')+2(l-l')=k+2l-1$. Thus every
element of $J^l_{k-1}(m')$, which now includes partitions with
arbitrary $m$, is an element of
$\cup^{l}_{l'=0}\cup^{k-1}_{k'=0}J^{l'}_{k';m'-2}\otimes
J^{l-l'}_{k-1-k'}$. Conversely, given any $j_L\in J^{l'}_{k';m'-2}$
and $j_R\in J^{l-l'}_{k-1-k'}$, the combined partition $j_L\otimes
j_R$ must be an element of $J^l_{k-1}(m')$, thus we have the
decomposition, \beq\label{decomp}
J^l_{k-1}(m')=\cup^{l}_{l'=0}\cup^{k-1}_{k'=0}J^{l'}_{k';m'-2}\otimes
J^{l-l'}_{k-1-k'}, \eeq where we have included all ways of
distributing $l$ internal points into two sets of $l'$ and $l-l'$
points; for a given $k'$, no partition exists beyond the range
$k'+l'-f'+1\leq m'-2\leq 2k'+2l'-2f'+1$ where $f'$ and $f-f'$ are
the numbers of faces in the two partitions respectively.

Now if we relabel the dummy variables, $j=i_{m'}, m_L=m'-2,
m_R=m-m'$ and note that $k'+l'-f'+1\leq m_L\leq 2k'+2l'-2f'+1$ and
$k-k'+l-l'+f-f'\leq m_R\leq 2(k-k')+2(l-l')-2(f-f')-1$ for given
$k',l',f'$, then by Eq.~(\ref{decomp}), $F$ splits into the left and
the right parts, \beqa
&&F=\frac{1}{l!}\sum_{\sigma_l}\sum^{n-1}_{j=4}[n,1,2,\widehat{j-1},j]\sum^{k-1}_{k'=0}\sum_{j_L^\alpha\in
J^{l'}_{k'}}\sum_{1<i_2\leq\ldots\leq i_{m_L+1}\leq j}
\int_{\{L\}}\prod_{e\in
E(j_L^\alpha)}[e](1,\ldots,j-1,j;\{A,B\}_L)\nn
 &&~~~
\times \sum_{j_R^\alpha\in J^{l-l'}_{k'-1-k'}}\sum_{j\leq
i_{m_L+3}\leq\ldots\leq i_{m_L+m_R+2}< n}\int_{\{R\}} \prod_{e\in
E(j_R^\alpha)}[e](j-1,j,\ldots,n\{A,B\}_R),\eeqa where $\sigma_l$
denotes the summation over all ways of distributing
$\{A,B\}_{\{l\}}$ into $\{A,B\}_L$ with $l'$ points, and $\{A,B\}_R$
with $l-l'$ points, for $l'=0,\ldots,l$, and the integral
$\int_{\{l\}}$ splits into $\int_{\{L\}}$ and $\int_{\{R\}}$
correspondingly.

To proceed, we note when $j=i_{m'}=i_{m'-1}$, a deformation
$\widehat{j-1}$ is needed in the factor $[n,1,2,\widehat{j-1},j]$,
which is an unwanted feature. However, due to the reversal symmetry
as in the tree-level case, one can shift the deformation from $j-1$
in the factor to the last leg $j$ in the left part, which is
$(j-1\,j)\cap (n\,1\,2)=I_j$ because the other vertex in the
preceding edge is $i_1=2$. In addition, note that $i_2-1$ needs to
be deformed when $i_2=i_1=2$, but, as the first external point in
$j_L^\alpha$, this is not a usual deformation inside the left part,
thus can only be achieved by a deformation of its first leg $1$,
$\widehat{1}=(1\,2)\cap(n\,j-1\,j)=\hat{1}_j$; similarly, the
deformation of $i_{m'+1}-1$ when $i_{m'+1}=i_{m'}=j$ is really a
deformation on the first leg of the right part, $j-1$,
$\widehat{j-1}=(j-1\,j)\cap(n\,1\,2)=I_j$. Then by the induction
assumption, the left and right parts are given by
$M_{j,k',l'}(\hat{1}_j,\ldots,j-1,I_j;\{A,B\}_L)$ and
$M_{n+2-j,k-1-k',l-l'}(I_j,j,\ldots,n;\{A,B\}_R)$, respectively,
\beqa\label{facallloop}
F&=&\frac{1}{l!}\sum_{\sigma_l}\sum^{n-1}_{j=4}[n,1,2,j-1,j]\Big[
\sum^{k-1}_{k'=0}M_{j,k',l'}(\hat{1}_j,\ldots,j-1,I_j;\{A,B\}_L)
 \nn
&& \kern+140pt\times
M_{n+2-j,k-1-k',l-l'}(I_j,j,\ldots,n;\{A,B\}_R)\Big]~, \eeqa which,
together with $M_{n-1,k,l}$, give exactly the factorization
contribution to $M_{n,k,l}$ in the recursion relations.

\begin{figure}
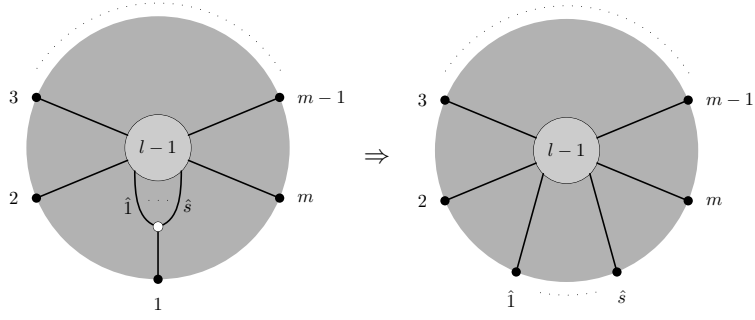
\centering
\includegraphicsbox[scale=0.7]{Loopfactsource.mps}~~$\Rightarrow$~~
\includegraphicsbox[scale=0.7]{Loopfactsource_b.mps}
\caption{Factorization of loop diagram of type $S$.}
\label{fig:loopdecompS}
\end{figure}
The remaining terms are those with $v=A_{l_0} (l_0=1,\ldots,l)$,
which all have a factor $[n,1,2,A_{l_0},\widehat{B_{l_0}}]$
 Fig(\ref{fig:loopdecompS}). As we
have shown in the one-loop MHV and NMHV cases, for each term one can
always redefine the integration variables to shift
$\widehat{B_{l_0}}$ back to $B_{l_0}$ without affecting anything
else. For a term to survive the integrations, $A_{l_0}$ in the
partition should also be connected to (at least one) other vertices
$v_i$ for $i=1,\ldots,s$. By pulling out the factor and considering
the rest as a new partition, one is essentially removing an external
point $i_1=2$ and an internal point $A_{l_0}$, and adding $s$
coinciding external points $v'_i=A_{l_0}$, which are connected to
$v_i$ for $i=1,\ldots,s$ respectively.

Note that the new partition, which is obviously also non-crossing,
has $l-1$ internal points, and $k+2l-1=k+1+2(l-1)$ edges which means
it belongs to the $k+1$ sector; the number of faces is $f-s+1\leq
f'\leq f$, and the number of external points is $m'=m+s-1$, which is
in the range $k+l-f+s,\ldots,2k+2l-2f+s$.
Denote the set of partitions found by pulling $s$ points to the
boundary as $J^{l-1}_{k+1}(s)$, this consists of diagrams which have
the last $s$ coinciding external points connected to $A_{l_0}$. Then
the full set of all partitions found by removing a leg connecting an
external vertex to an internal vertex, $S^l_k$ is the union of
$J^{l-1}_{k+1}(s)$ for all possible $s$. Conversely taking any
element of $S^l_k$ and joining together the set of legs connected to
$A_{l_0}$ we find an element of $J^{l-1}_{k+1}(s)$ for some $s$.
Thus \beq S^{l}_{k}=\cup_{s}J^{l-1}_{k+1}(s)~. \eeq

In addition to the summation over $j^\alpha\in J^{l-1}_{k+1}(s)$, we
also sum over the previous external points, $1<i_2\leq\ldots\leq
i_{m'-s+1}<n$, while constraining the added, last $s$ external
points to be fixed at $A_{l_0}$, i.e.
$i_{m'-s+2}=\ldots=i_{m'+1}=A_{l_0}$, for all possible $s$, \beqa
\label{sourceallloop}
&&S=\frac{1}{l!}\sum^l_{l_0=1}\int_{\{l\}/l_0}\int_{l_0}[n,1,2,A_{l_0},B_{l_0}]\sum_{s}\sum_{j^\alpha\in
J^{l-1}_{k+1}(s)}\sum_{1<i_2\leq\ldots\leq i_{m'-s+1}<n}\nn
&&\prod_{e\in
E(j^\alpha)}(\pm)[n,\widehat{v_1(e)-1},v_1(e),\widehat{v_2(e)-1},v_2(e)]|_{i_{m'-s+2}=\ldots=i_{m'+1}=A_{l_0}},\eeqa
where the contribution is denoted as $S$. Since we have included all
permutations of internal points, any internal point $A_{l_0}$ can be
connected to $i_1=2$, thus there is a summation over $l_0$.

Now if we relax the constraints and generally sum over $2\leq
i_2\leq\ldots\leq i_{m'+1}\leq n+2$ with $n+1=B_{l_0}$ and
$n+2=A_{l_0}$, then by splitting the last coinciding $s$ points of
any partition in $J^{l-1}_{k+1}(s)\subseteq S^l_k$, one obtains a
partition in $J^{l-1}_{k+1}$; conversely, by joining the last $s$
points of any partition in $J^{l-1}_{k+1}$ for any possible $s$, we
obtain either a partition in $J^{l-1}_{k+1}(s)\subseteq S^l_k$, or a
partition with multiple edges connecting the same pair of points,
which vanishes immediately. Denoting the set of these bubble
diagrams as $B^l_k$, we have, \beq S^l_k\cup B^l_k=J^{l-1}_{k+1}.
\eeq

By the induction assumption, and note that there is no contribution
from $B^l_k$, then the summations and the integrations
$\int_{\{l\}/l_0}$, as well as a factor $1/(l-1)!$, give
$M_{n+2,k+1,l-1}$, \beqa
&&\frac{1}{(l-1)!}\int_{\{l\}/{l_0}}\sum_{j^\alpha\in
J^{l-1}_{k+1}}\sum_{2\leq i_2\leq\ldots\leq i_{m'+1}\leq
n+2}\prod_{e\in
E(j^\alpha)}(\pm)[n,\widehat{v_1(e)-1},v_1(e),\widehat{v_2(e)-1},v_2(e)]\nn
&&=M_{n+2,k+1,l-1}(\hat{1}_{A_{l_0}B_{l_0}},\ldots,n,\hat{B_{l_0}},A_{l_0};\{A,B\}_{\{l\}/{l_0}}),\eeqa
where one needs a deformation on $1$ because the deformation for
$i_2$ when $i_2=i_1=2$ is only achieved by always using
$\widehat{1}=(1\,2)\cap
(n,\,A_{l_0}\,B_{l_0})=\hat{1}_{A_{l_0}B_{l_0}}$; also the
deformation $\widehat{B_{l_0}}=(A_{l_0}\,B_{l_0})\cap(n\,1\,2)$ is
not an usual deformation in $M_{n+2,k+1,l-1}$, thus it is only
achieved by always using $\hat{B_{l_0}}$, and by the same argument
before, all other deformed $B_{l_0}$ can be defined by the same
deformation acting on $\hat{B_{l_0}}$.

By relaxing the constraints, we have included some unwanted terms.
Since Eq.~(\ref{sourceallloop}) already has all terms with at least
one factor ($s\geq 1$) containing both $A_{l_0}$ and $B_{l_0}$ (in
addition to the prefactor $[n,1,2,A_{l_0},B_{l_0}]$), as we have
shown, all unwanted terms vanish after the integration $\int_{l_0}$.
Thus we can indeed write the contribution $S$ as an integration of
$M_{n+2,k+1,l-1}$, \beq
S=\frac{1}{l}\sum^l_{l_0=1}\int_{l_0}[n,1,2,A_{l_0},B_{l_0}]M_{n+2,k+1,l-1}(\hat{1}_{A_{l_0}B_{l_0}},\ldots,n,A_{l_0},\hat{B}_{l_0};\{A,B\}_{\{l\}/{l_0}}),\eeq
where we have exchanged $A_{l_0}$ and $\hat{B}_{l_0}$ because in the
surviving terms, either order gives the same result. Finally, by
combining $F$ from Eq.~(\ref{facallloop}) and the source term $S$
from Eq.~(\ref{sourceallloop}) together, \beq
M_{n,k,l}(1,\ldots,n;\{A,B\}_{l})=M_{n-1,k,l}(2,\ldots,n;\{A,B\}_{l})+F+S,\eeq
we have seen that Eq.~(\ref{eq:allloop}) indeed gives an explicit
solution to the recursion relations to all loops. This provides
strong evidence for the all-loop MHV vertex expansion
Eq.~(\ref{eq:allloop}).

We stress again that the general formula Eq.~(\ref{eq:allloop}) is
completely cyclically invariant and thus it can generate different
forms of all-loop integrands by choosing different reference
twistors. By considering generalized BCFW relations with different
shifts one can see that the expression Eq.~(\ref{eq:allloop}) is
valid for a choice of the reference twistor equal to any of the
external twistors. However, this does not prove that the formula is
completely independent of the choice of such a reference twistor. It
should be possible to prove this by considering all-line shifts as
in the standard case.
\begin{figure}
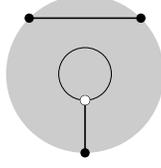
\centering
\includegraphicsbox[scale=0.7]{One_loop_NMHV_tadpole.mps}
\caption{One-loop NMHV tadpole partition.}
\label{fig:One_loop_NMHV_tadpole}
\end{figure}

Just as we have shown that the MHV diagrams corresponding to
the graphs in $J^{l}_{k}$ satisfy a recursion relation, one can similarly derive a
recursion relation for the number of graphs. To be specific, let us
consider the class of non-crossing graphs described above: with $m$
external points, $l$ internal points, and $k+2l$ edges. We define  $C^{l}_{k;m}$
to the number of non-crossing partitions or dual graphs, restricted
 to graphs with distinct external points, i.e. they all have exactly
three adjacent edges, but relaxing the definition to include graphs with internal faces with one or two edges, i.e. bubbles or tadpoles attached to internal vertices, for example
$C^1_{1;3}$ would now include Fig.(\ref{fig:One_loop_NMHV_tadpole}) in addition
to Fig.(\ref{fig:One_loop_NMHV_Partitions}) (a). $C^{l}_{k;m}$ satisfies the recursion relation
\beqa
C^{l}_{k;m}&=&\sum_{l'=0}^l\sum_{k'=0}^{k-1}\sum_{m'=2}^m C^{l'}_{k';m'-2} C^{l-l'}_{k-k'-1;m-m'}+\sum_{s=1}^{s_{max}} C^{l-1}_{k+1;m+s-1}
\eeqa
where $s_{max}=1$ if $C^{l}_{k;m}$ has no internal faces, $s_{max}=2$ if $C^{l}_{k;m}$ has
one internal face, and  $s_{max}=2+f_{max}$ for $f_{max}>1$ where $f_{max}$ is the
maximum number of allowed number of faces in $C^{l}_{k;m}$
which is 0 for $l=0$ and $k+l-m+1$ for $l< 0$. As a boundary condition, we define $C^{0}_{0;0}=1$.

\section{Conclusions and Outlook}

In this work we have shown that the expressions for the tree-level
amplitudes and loop integrands following from the momentum twistor
space MHV vertex expansion satisfy the ABCCT recursion relations.
This provides strong evidence for the validity of the MHV vertex
expansion to all loop order and the independence of the expressions
on the choice of reference twistor.
\footnote{We have shown that the MHV expansion is valid for a choice of the
reference twistor equal to any of the external twistors.}
Correspondingly, the expressions
from the MHV expansion provide manifestly cyclicly invariant solutions
of the ABCCT recursion relations.

In BCFW and ABCCT recursion relations, the many important properties
of tree amplitudes and loop integrands, such as cyclic-invariance,
absence of spurious poles, correspond to a set of highly non-trivial
relations between rational functions~\cite{ArkaniHamed:2009dn}. At
tree-level, these relations can be understood as arising from the
global residue theorem applied to the Grassmannian
integral~\cite{ArkaniHamed:2009dn, Broedel:2010rr}, and
generalizations to loop-level have been suggested
in~\cite{ArkaniHamed:2010kv}. It would be interesting to see how
those relations, or the residue theorem, arise from the explicit
solution which unifies different the forms of loop integrands.
Furthermore, it is possible to obtain the general local form of
integrands for all numbers of legs and all loop orders, generalizing
results of~\cite{ArkaniHamed:2010kv}.

Of course one is ultimately interested in the integrated expressions
and so a regulator for the IR divergences must be introduced. A
convenient choice, which can be implemented in momentum twistor
space \cite{Hodges:2010kq, Mason:2010pg}, is a mass-regulator which
corresponds to higgsing the theory e.g. \cite{Alday:2007hr,
Alday:2009zm} (see \cite{Alday:2010jz, Drummond:2010mb} for recent
use of this regulator in explicit calculations). While this
regulator is natural from the dual conformal point of view, the full
Yangian symmetry, in particular the usual conformal symmetry, is
still not well-understood in the presence of IR divergences and
na\"ively it becomes anomalous. At tree-level it was shown that the
superconformal and dual superconformal symmetries, properly
understood \cite{Bargheer:2009qu}, \cite{Sever:2009aa}, could be
used to uniquely fix the tree level amplitudes
\cite{Bargheer:2009qu}. This was equivalent to using the na\"ive
symmetry generators and accounting for the collinear or soft
behavior of the amplitudes \cite{Korchemsky:2009hm}. At loop level
it was further shown how to deform the symmetry generators, to
account for the anomalous contributions arising from regulator,
acting on the full one-loop amplitudes \cite{Beisert:2010gn}. In
that work the amplitudes were all defined using the version of
dimensional reduction commonly used in explicit calculations of
amplitudes. A related prescription for the symmetry generators,
based on an all-loop generalization of the CSW prescription combined
with a novel regulator was proposed in \cite{Sever:2009aa}. Since we
have written down all-loop integrands in a concise form, it would be
interesting to see if the symmetries, again defined appropriately,
can be used to determine these integrands. To extend this to the
full all-loop amplitudes it may be useful to reconsider these
earlier calculations in the momentum twistor space formulation and
with the mass regulator. In a perhaps related direction,
differential operators which relate integrals at different loop
orders were recently found in momentum twistor space
\cite{Drummond:2010cz}.

In work closely related to the development of the momentum twistor
MHV expansion, Mason and Skinner \cite{Mason:2010yk} showed that the
correlations of the Wilson loop in momentum twistor space lead
exactly to the MHV expansion for scattering amplitudes but with the
diagrams being the planar duals of the MHV diagrams. Another,
presumably equivalent or related, proposal for a supersymmetric
Wilson loop \cite{CaronHuot:2010ek}, building on the considerations
of \cite{Eden:2010zz, Eden:2010ce},  has been shown to satisfy
recursion relations equivalent to those of ABCCT at tree and loop
level. Relatedly, Brandhuber et al \cite{Brandhuber:2010mi}
presented a set of dual momentum space rules, interpreted as dual
momentum space Wilson loop diagrams. These rules are equivalent to
the ordinary MHV rules and simply correspond to their dual graph
representation. In our work we have introduced non-crossing
partitions which are in one-to-one correspondence with the dual MHV
graphs. It may be fruitful to understand exactly the relation
between all these prescriptions. In particular, the explicit formula
we obtained should naturally follow from a dual graph expansion of
the Wilson loops.

In demonstrating that the MHV expansion satisfies
 the ABCCT relations we have made use of the freedom to choose the
reference twistor. However, while it is almost
certainly true that the expressions are independent of this choice
it would be nice to prove it. Our result is valid for a choice of 
the reference twistor equal to any of the external twistors, since it
can be derived from recursion relations by shifting any external
twistor. Thus this provides some evidence for the arbitrary nature
of the reference twistor. In a very recent paper \cite{Bullimore:2010dz}, Bullimore
adopted a complementary method to show the MHV expansion satisfies
recursion relations derived from the momentum-twistor version of the
all-line shift \cite{Risager:2005vk, Elvang:2008vz}, thus it is at
least valid for all reference twistors of the form
$\mathcal{Z}_{\ast}=(0,\iota^{\dot{\alpha}},0)$.

It may also be possible to extend these results to wider class of
observables such as form factors. The string dual of these
observables is also described by an integrable model and their
strong coupling value is calculable by a set of equations of the
Y-system form \cite{Maldacena:2010kp}. At weak coupling there is
evidence \cite{Bork:2010wf} of dual conformal symmetry or at least a
residue thereof and so they may be expressible in terms of Yangian
invariants. It would be interesting to investigate whether recursion
relations for these quantities can be found.

Finally, we emphasize that rewriting the MHV expansion in terms of
non-crossing partitions has the advantage of systematically
organizing all planar MHV diagrams, and yielding more explicit
results. As a byproduct, our expressions for the MHV expansion,
combined with the ABCCT relations, provide a simple recursive formula
for the number of generalized non-crossing partitions or
equivalently a particular class of dual planar graphs.


\vspace{1cm} \noindent {\bf Acknowledgements:} We would like to
thank G. S. Vartanov for useful conversations and kindly informing
us of \cite{Bork:2010wf}. We are also grateful to M. Bullimore for
several insightful comments and questions.

\appendix

\section{Non-crossing partitions and dual diagrams}
\label{app:diagrams} 
\begin{figure}
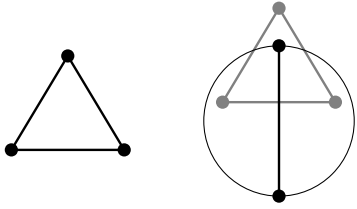
\centering
\includegraphicsbox[scale=1]{TreeNMHVgraph.mps}~~~~~~~~\includegraphicsbox[scale=1]{TreeNMHVdgraph.mps}
\caption{Graph and dual graph for the NMHV diagram.The uppermost
point of the graph is the added vertex upon which the external legs
end.} \label{fig:TreeNMHVgraph}
\end{figure}

The non-crossing partitions we have introduced
are obviously closely related to the dual graphs of the MHV vertex
diagrams composing the amplitude (which have appeared in this
context recently in  \cite{Mason:2010yk,
Brandhuber:2010mi}). Here we attempt to make this
relationship clearer and in doing so argue for the validity of our
expression Eq. (\ref{eq:allloop}). Starting with the simplest tree-level
diagram, Fig.(\ref{fig:FigTreeNMHV}), we write the MHV vertex as a
graph vertex. First we collapse each group of external legs into a
single leg ending on a new vertex, and we furthermore identify all
these new vertices, see Fig.(\ref{fig:TreeNMHVgraph}). In general, this
produces a (multi)graph on the sphere with faces
corresponding to momentum regions. Of course, graphs on a sphere are
in one-to-one correspondence with planar graphs (where one of the
faces corresponds to the exterior region). For every planar graph
one can construct the geometric dual graph, for example
Fig.(\ref{fig:TreeNMHVgraph}), which is seen, in this case, to
correspond to the non-crossing partition with the external vertices
connected by edges.

\begin{figure}
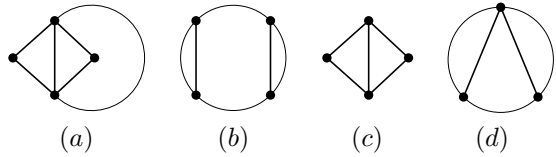
\centering
\begin{eqnarray}
\begin{array}{cccc}
\includegraphicsbox[scale=0.7]{TreeNNMHVgraph.mps}~~ &
\includegraphicsbox[scale=0.7]{TreeNNMHVdgraph.mps}~~ &
\includegraphicsbox[scale=0.7]{TreeNNMHVgraphdegen.mps}~~ &
\includegraphicsbox[scale=0.7]{TreeNNMHVdgraphdegen.mps} \nn
  (a)~~&  (b)~~& (c)~~& (d)~~
\end{array}
\end{eqnarray}
\caption{Tree N$^2$MHV graph (a), the dual graph (b), the degenerate
N$^2$MHV graph (c) and the dual of the degenerate graph. }
\label{fig:FigTreeNNMHVgraph}
\end{figure}

While for more complicated tree diagrams the resulting graphs are
more involved one can always construct a dual graph in this
fashion. Furthermore, the dual of each graph is unique. These planar
graphs are always such that each edge ends on two distinct vertices
(no pseudographs), and in fact each face is bounded by exactly three
edges. This corresponds to assuming that for each location in the original MHV
diagram where external legs can occur they do. The case where there
are no external legs, for example if $i_2=i_3$ in
Fig.(\ref{fig:FigTreeNNMHV}), is a degenerate case which corresponds
to two vertices becoming coincident in the dual graph. For example
in Fig.(\ref{fig:FigTreeNNMHVgraph}) we show the graph corresponding
to the N$^2$MHV diagram, the dual graph, the degenerate graph for
$i_2=i_3$ and its dual.

Furthermore, before identifying the end vertices there are no
cycles, \footnote{By cycle we mean a closed path with no repeated
edges or vertices except the starting/ending vertex.} thus after the
identification every cycle must include the added vertex
corresponding to the identified ends of all external legs. That is
to say, the graphs are such that every cycle has at least one vertex
in common. For the dual diagrams this implies that every vertex is
connected to at least three edges and there is one face (in our
embedding this is the exterior region) such that every cut set
\footnote{By a cut set we mean the set of edges dual to the edges of
a cycle. Removing those edges results in a partition of the set of
vertices, which defines a cut. Alternatively, the cut set is the set
of all edges whose end points are in different partitions of a given
cut.}
\begin{figure}
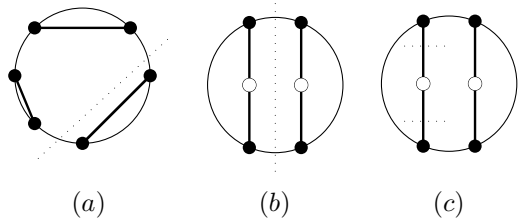
\centering
\begin{eqnarray}
\begin{array}{ccc}
\includegraphicsbox[scale=1]{Treecutdgraph.mps}~~&\includegraphicsbox[scale=1]{Loopcutdgraph_a.mps}~~&\includegraphicsbox[scale=1]{Loopcutdgraph_b.mps}\nn
 (a)~~&  (b)~~& (c)
 \end{array}
\end{eqnarray}
\caption{(a) Cut for a  dual tree graph (b) Boundary cut for a dual graph (c) Loop cut for a dual graph}
\label{fig:Cutgraphs}
\end{figure}
 must contain two edges
of the boundary of that face e.g. Fig.(\ref{fig:Cutgraphs}) (a). In the degenerate graphs vertices can
come together, in which case one could have more than three edges
attached to a vertex.

\begin{figure}
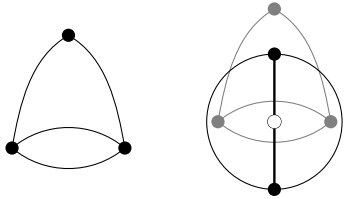
\centering
\includegraphicsbox[scale=1]{OneloopMHVgraph.mps}~~~~~~~~\includegraphicsbox[scale=1]{OneloopMHVdgraph.mps}
\caption{Graph and dual graph for the one-loop MHV diagram.}
\label{fig:OneloopMHVgraph}
\end{figure}
Now we consider MHV diagrams with loops, the simplest example of
which, the one loop MHV diagram, is shown in
Fig.(\ref{fig:OneloopMHVgraph}). As can be seen even in this
simplest case, for the MHV diagrams we must include faces
corresponding to loops which in general  can have more than three boundary
edges; we thus have two classes of faces. As we do not consider
tadpoles, as they give zero after fermionic integration,
those faces corresponding to loops must have at least two
edges each.
In fact, graphs where an internal
face has only two edges but shares one of its edges with another
loop will also give zero contribution to the MHV expansion after
the fermionic integration. This is the case, for example,
in the two loop MHV diagrams Fig.(\ref{fig:Two_loop_MHV_diagrams}) (b).
Contrary to the tree level graphs, there
are now, of course, completely disjoint cycles.

For the corresponding dual diagrams we thus have two types of
vertices, which we represent with black dots for vertices dual to
faces corresponding to external momenta and white dots for vertices
dual to faces corresponding to loop momenta. Black dots must always
have three adjacent edges and white dots at least two edges. In
fact, for the purposes of integrands, white dots must have two edges
each, that is they cannot each have only two edges but share an
edge. Every face must have a boundary consisting of at least three
edges. Finally, there is a face (in our representations this will be
the exterior face) such that the cut sets of  cuts that result in
two partitions both consisting of mixes of white and black vertices
must contain elements of the boundary of this face e.g.
Fig.(\ref{fig:Cutgraphs}) (b). The cut sets of cuts which result in
partitions one of which contains only white vertices contain no
elements of the boundary of this face e.g. Fig.(\ref{fig:Cutgraphs})
(c). These cuts correspond to the cycles introduced by the loops.

The partitions described earlier can be found by eliminating the
boundary of the preferred, or exterior face. This gives us a
one-to-one mapping between MHV diagrams, their graphs, their duals
and the corresponding partitions.

\bibliographystyle{nb}
\bibliography{All_loop}

\end{document}